\documentclass[12pt,preprint]{aastex}
\def\R23{\mbox{$\rm R_{23}$}}

\def\arcsec{\hbox{$^{\prime\prime}$}}

\def\kmsmpc{km s$^{-1}$ Mpc$^{-1}$}

\def\Hb{\mbox{${\rm H}{\beta}$}}
\def\Ha{\mbox{${\rm H}{\alpha}$}}
\def\OIII{\mbox{${\rm [O\,III]\,}{\lambda\lambda\,4959,5007}$}}

\def\OIIIa{\mbox{${\rm [O\,III]\,}{\lambda\,5007}$}}
\def\OIIIb{\mbox{${\rm [O\,III]\,}{\lambda\,4959}$}}
\def\OII{\mbox{${\rm [O\,II]\,}{\lambda\,3727}$}}

\def\NII{\mbox{${\rm [N\,II]\,}{\lambda\,6584}$}}

\shorttitle{[O/H], SFRs and Dust at Intermediate z}
\shortauthors{Maier, C. et al.}

\begin{document}


\title{Near-Infrared Spectroscopy of $0.4<z<1.0$ CFRS  Galaxies: Oxygen Abundances, SFRs and Dust\footnotemark[0]}

\author{C. Maier\altaffilmark{1}}
%\affil{Department of Physics, Swiss Federal Institute of Technology (ETH Z\"urich), ETH H\"onggerberg, CH-8093, Z\"urich, Switzerland}
\email{chmaier@phys.ethz.ch}

%\and

\author{S.J. Lilly\altaffilmark{1}, M. Carollo\altaffilmark{1},
  A.  Stockton\altaffilmark{2}, and M. Brodwin\altaffilmark{3}}

\footnotetext{Based on observations
  obtained at the ESO VLT, Paranal, Chile; ESO programs 070.B-0751,
and 072.B-0496}
\altaffiltext{1}{Department of Physics, Swiss Federal Institute of Technology (ETH Z\"urich), ETH H\"onggerberg, CH-8093, Z\"urich, Switzerland}
\altaffiltext{2}{Institute for Astronomy, University of Hawaii, 2680
  Woodlawn Drive, Honolulu, HI 96822}
\altaffiltext{3}{Jet Propulsion Laboratory, Caltech, 4800 Oak Grove
  Drive, Pasadena, California 91109}

\begin{abstract}
   Using new J-band VLT-ISAAC and Keck-NIRSPEC spectroscopy, we have
  measured \Ha\, and \NII\, line fluxes for  $0.47<z<0.92$ CFRS galaxies
  which have \OII, \Hb\, and \OIIIa\, line fluxes available from 
 optical spectroscopy,
 to investigate how the
  properties of the star forming gas in galaxies evolve with redshift.
  We derive the extinction and oxygen abundances for the  sample using a method  based on
  a set of ionisation parameter and oxygen abundance diagnostics, 
%developed by  \citet{kewdop02pap},
  simultaneously fitting the [OII], \Hb,
  [OIII], \Ha, and [NII] line fluxes.  The individual reddening measurements allow us to
  accurately correct the \Ha\,-based star formation rate (SFR)  estimates for
  extinction. 
Our most salient conclusions are:
a) in all 30  CFRS galaxies the  source of gas ionisation is not due
to AGN activity;
b) we find a range of $0<A_{V}<3$, suggesting  that it
  is important to determine the extinction for every
  single galaxy in order to reliably measure SFRs and oxygen abundances in high
  redshift galaxies;
c) high values of \NII/\Ha\, $>0.1$ for most (but not all) of the  CFRS galaxies indicate
that they lie on the high-metallicity branch of the \R23\, calibration;
d) about one third of the $0.47<z<0.92$ CFRS
  galaxies in our sample have lower metallicities than local galaxies with similar
  luminosities and star formation rates;
% and the average change in
%  metallicity is 0.3\,dex between the CFRS objects and local galaxies of
%  similar luminosities;
e) comparison with a chemical evolution model indicates that these low
  metallicity galaxies
 are unlikely to be the progenitors of metal-poor dwarf galaxies at
 $z\sim0$.
\end{abstract}


%######################################################################################

\section{Introduction}

The starlight that is seen at any particular redshift must be associated with
the production of heavy elements which can then be seen (in stellar
atmospheres or in the gas phase) at all later epochs.  Monitoring how the
chemical content of galaxies changes with cosmic time is thus important not
only as a {\sl per se} diagnostic of galaxy evolution, but also to constrain
the history of the global star formation activity in the universe.

For many years, there have been extensive studies of the chemical composition
of gas in the high redshift Universe through studies of quasar absorption line
systems
\citep[see the review by][and references therein]{pettini03}.
More recently, there has been growing attention to measurements of the
metallicities of the star-forming gas observed in distant galaxies at
significant look-back times
\citep{calilly01pap,hammer01,lilly03,kob03,maier04,mehlert02,pettini02}.
In the main, these measurements have focussed on the oxygen abundance,
[O/H], as computed from a number of empirically calibrated metallicity
estimators based on the relative strengths of strong emission lines. The most
popular of these estimators has  been the $\rm R_{23}$ parameter
introduced by Pagel et al.\ (1979), which is defined as $\rm R_{23}= ({\rm
  [O\,II]\,}{\lambda\,3727} + {\rm [O\,III]\,}{\lambda\lambda\,4959,5007})/
{\rm H}{\beta}$.

A main reason for the success of $\rm R_{23}$ as a gas metallicity estimator
at intermediate redshifts is that it is relatively easy to measure, as the
observed wavelengths of the $\rm {[O\,II]\,]}{\lambda\,3727}$, ${\rm
  [O\,III]\,}{\lambda\lambda\,4959,5007}$ and ${\rm H}{\beta}$ lines remain,
at $z<1$, in the optical window.  Indeed, the use of $\rm R_{23}$ has
provided key information, albeit coarse, on the gas metallicities of several
dozens of intermediate redshift galaxies, and thus opened the possibility of
investigating, in a statistically meaningful way, the chemical evolution with
cosmic time of the normal galaxy population.  
Specifically, after initial results for a subsample of 15 galaxies were
published in CL01, LCS03 provided gas
oxygen abundances  based on $\rm R_{23}$ for a total sample of 66 $0.47<z<0.92$
CFRS galaxies.

There are however some limitations in all work based on the $\rm R_{23}$
estimates.  First, it is well known that $\rm R_{23}$ is degenerate with
metallicity, as low \R23\, values can be associated with either high or very
low [O/H] values. Furthermore, $\rm R_{23}$ is significantly affected
by dust extinction.  
In CL01 and LCS03 it was \emph{assumed} that the CFRS galaxies lie on the
high-metallicity branch of the \R23\,-metallicity relation, and  the
spectra were dereddened assuming a uniform $E_{B-V}=0.3\pm 0.15$, i.e., the average for
galaxies of the Nearby Field Galaxy Sample \citep{jansen}.  Eliminating these
sources of potential error requires a  determination of the reddening as
well as breaking the \R23 degeneracy, for each galaxy
  individually. This can be achieved with knowledge of the \Ha\, and \NII\,
lines fluxes, which, at intermediate redshifts, are shifted into the
near-infrared (NIR)
wavelength window. Specifically, the reddening by dust can be estimated by
means of  Balmer lines ratios, e.g., the \Ha/\Hb\, ratio, and the degeneracy in
\R23 can be broken by measuring the NII/\Ha\, ratio.

In this paper we present new measurements of \Ha\, and \NII\, for 28 of the
$0.47<z<0.92$ CFRS galaxies investigated in LCS03 supplemented by
CFHT optical spectra to derive \Ha\, and \NII\, for two galaxies at
$z=0.472$ and $z=0.479$.
We use these new measurements to derive accurate estimates for the gas oxygen
abundance, dust extinction and star formation rate (SFR) in the sample galaxies.
The paper is structured as follows: In Sect. 2 we describe the new near
infrared
data, acquired with VLT/ISAAC and Keck/NIRSPEC, and the data reduction
procedure.  In Sect. 3 we describe the derivation of the oxygen abundances.
These were obtained through a simultaneous fit of the \OII, \Hb, \OIIIa, \Ha,
and \NII\, lines in terms of extinction parameter $A_{V}$, ionisation
parameter $q$, and [O/H], using the set of ionisation parameter and oxygen
abundances developed by \citet[][hereafter KD02]{kewdop02pap} .  This procedure provides
metallicity estimates which should be more robust than those obtained from
individual line ratios (as done, e.g., by
\citet{kob03}, \citet{kobkew04}, \citet{shapley04}, and in our own CL01 and
LCS03).  In Sect. 4 we present the results of our analysis, i.e., the
extinction, SFRs, and oxygen abundances of the 30 CFRS galaxies. We then compare these measurements with the properties of galaxies in the
local universe, and discuss a scenario for the chemical evolution of
galaxies at $0<z<1$.
Finally, in Sect. 5 we present our conclusions.

A ``concordance'' cosmology with $\rm{H}_{0}=70$ \kmsmpc, $\Omega_{0}=0.3$,
$\Omega_{\Lambda}=0.7$ is used throughout this paper. 
Note that, unless otherwise specified, the terms {\it metallicity} and {\it
  abundance} are used to indicate {\it oxygen abundance} of the
 line emitting gas throughout this
paper.

%#######################################################################################

\section{Near-infrared observations, data reduction, and line flux measurements} 

The 30 $0.47<z<0.92$ CFRS galaxies investigated in this paper are extracted
from the 66 objects sample presented in LCS03. They form an essentially
random sub-sample. All the galaxies have
therefore \OIIIa, \Hb, and \OII\, lines measured with the CFHT.  The galaxies
have absolute B magnitudes $\rm{M}_{\rm{B,AB}}<-19.5$.

For two objects at
$z=0.472$ and $z=0.479$, \Ha\, and \NII\, measurements could already be obtained from the optical CFHT
spectra presented in LCS03. For the remaining 28 galaxies discussed in
this paper, new \Ha\, and \NII\, measurements were obtained from  VLT and Keck NIR
spectroscopy.
For easy reference, we report in Table \ref{infoobj} the
photometry, line fluxes (including the new \Ha\, and \NII), and other information we have collected for the 30 CFRS galaxies.

\subsection{The VLT-ISAAC spectra}

Near-infrared spectroscopy for 17 $0.5<z<0.92$ CFRS galaxies of LCS03 was
obtained with the ISAAC spectrograph at the VLT in order to measure their
\Ha\,  and 
\NII\, line fluxes.
The observations were carried out in October 2002 (Program 070.B-0751A,
hereafter P70), and November/December 2003 plus September/October 2004
(Program 072.B-0496A, hereafter P72). The medium resolution grism was used
with the Short-Wavelength channel equipped with a $1024 \times 1024$ pixel
Hawaii array. The pixel scale is 0.147\arcsec\, per pixel.
We used a slit of 2\arcsec\, width, which results in a nominal resolution of
$\rm{R} \sim 1500$.  

We used two different filters, i.e., the SZ and the J filter, in order
to select the 5th and 4th grating order for measuring the \Ha\, and
\NII\, lines of galaxies at $0.5<z<0.65$ and $0.7<z<0.92$, respectively.
The covered wavelength range was 59\,nm and 46\,nm when using the J and the SZ
filter, respectively.  The corresponding pixel scales were $0.58$\AA/pixel and
$0.45$\AA/pixel.

The individual integration times varied between 1200\,s and 4800\,s.  During
the observations the telescope was nodded between two positions
$\sim10$\arcsec\, apart along the slit in P70, and between eight positions
$\sim5$\arcsec\, apart in P72.  Dark frames, flat-fields and (Xe+Ar) arc lamp
spectra were taken with the same filter, central wavelength and slit width for
each of the targets observed during the night.  The conditions were
photometric during these observations (see Table~\ref{ObsISAAC} for further
details).

\subsubsection{Processing of the 2D galaxy spectra}

$1024\times800$ pixel frames were extracted from the original $1024\times1024$
pixel raw frames.  A flat field was obtained by normalizing the difference
between a lamp-on and a lamp-off image.  The individual science frames were
subtracted of dark, and flat-fielded.  Cosmic rays were removed 
in every single exposure using the IRAF routine \emph{cosmicrays} .
Wavelength calibration and a geometrical distortion correction were performed
using arc spectra of xenon and argon. A 3rd order Legendre polynomial provided an
adequate fit to the dispersion relation.  Typical relative uncertainties of
the dispersion relation were of order $\sim 1/10$ of a pixel, corresponding to
$\sim 0.06$\AA.
The background was determined by applying the IRAF procedure \emph{median} to
each individual 2D spectra using a sliding rectangular window of dimension
$1\times201$ pixels in the x and y direction, respectively. This background
was subtracted from each science frame.  

For each   galaxy, a median sky was obtained by masking out the galaxy spectrum in each  of the (two or eight for P70 and P72, respectively) individual dithered
  frames, and applying a median algorithm to the stack of masked frames.  The
  median sky was then subtracted from each individual dithered image. The
  final 2D spectrum was obtained by aligning and averaging these
  sky-subtracted frames.

%*************

\subsubsection{Spectrophotometric calibration}
\label{speccal}

Calibration of spectroscopic data in the NIR 
is symplified by  the use of stars hotter than B4. These have relatively featureless
spectra and allow one to correct the telluric features of the Earth's
Atmosphere and to calibrate fluxes in the NIR wavelength window.  We used
six telluric standards from the Hipparcos
  catalogue to calibrate our VLT spectra (see Table~\ref{ObsISAAC}). The
  six stars were observed under photometric conditions.  These stars
  have measured V band magnitudes, V-I colours, spectral types, and JHK band
  magnitudes.  This leads to uncertainties of about 10\% in the absolute flux
  calibration.

 After bias subtraction, flat fielding, wavelength calibration and geometric
distortion correction using IRAF routines, the standards 1D spectra were extracted from the
resulting 2D frames, the sensitivity response as a function of
wavelength was determined using a 3rd order Legendre function, and the
the flux calibration was done, after correcting the spectra for (small) air-mass effects.

%************************

\subsubsection{Extraction of the one-dimensional galaxy spectra}

\begin{figure}[h!]
\plotone{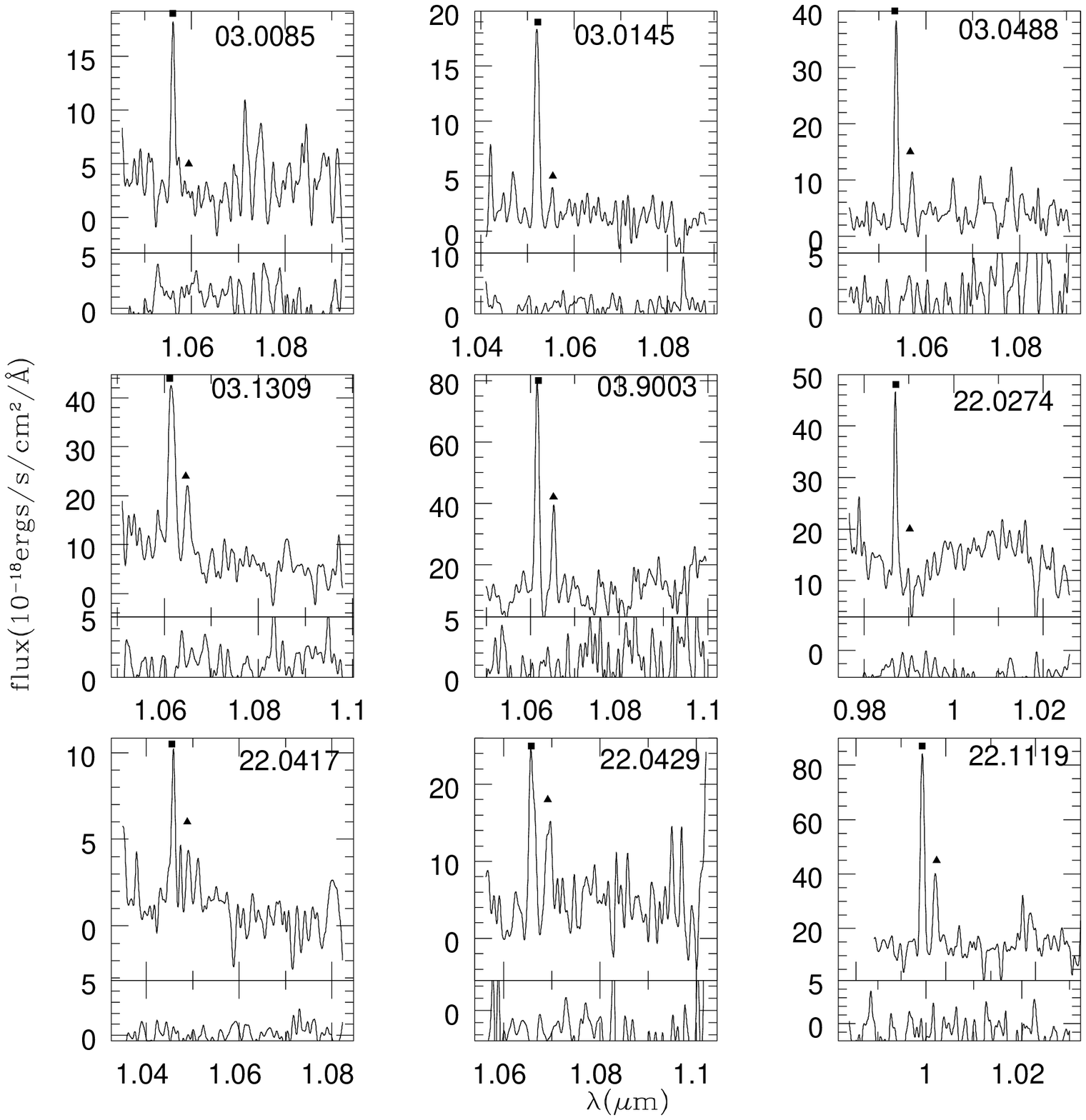}
\caption
{\label{specSZ} \footnotesize ISAAC spectra using the SZ filter for the
  5th grating order in order to get \Ha\, and  \NII\, for galaxies at $0.5<z<0.65$.
The spectra are   smoothed with a Gaussian filter, as described in the
text. In each figure, the filled square and  the filled triangle show
  the position of \Ha\, and  \NII, respectively. 
Almost all galaxies show a \NII/\Ha\, ratio greater than 0.1, as
  expected for galaxies that lie on the high-metallicity branch of the
  \R23\, calibration.
}
\end{figure}

\begin{figure}[h!]
\plotone{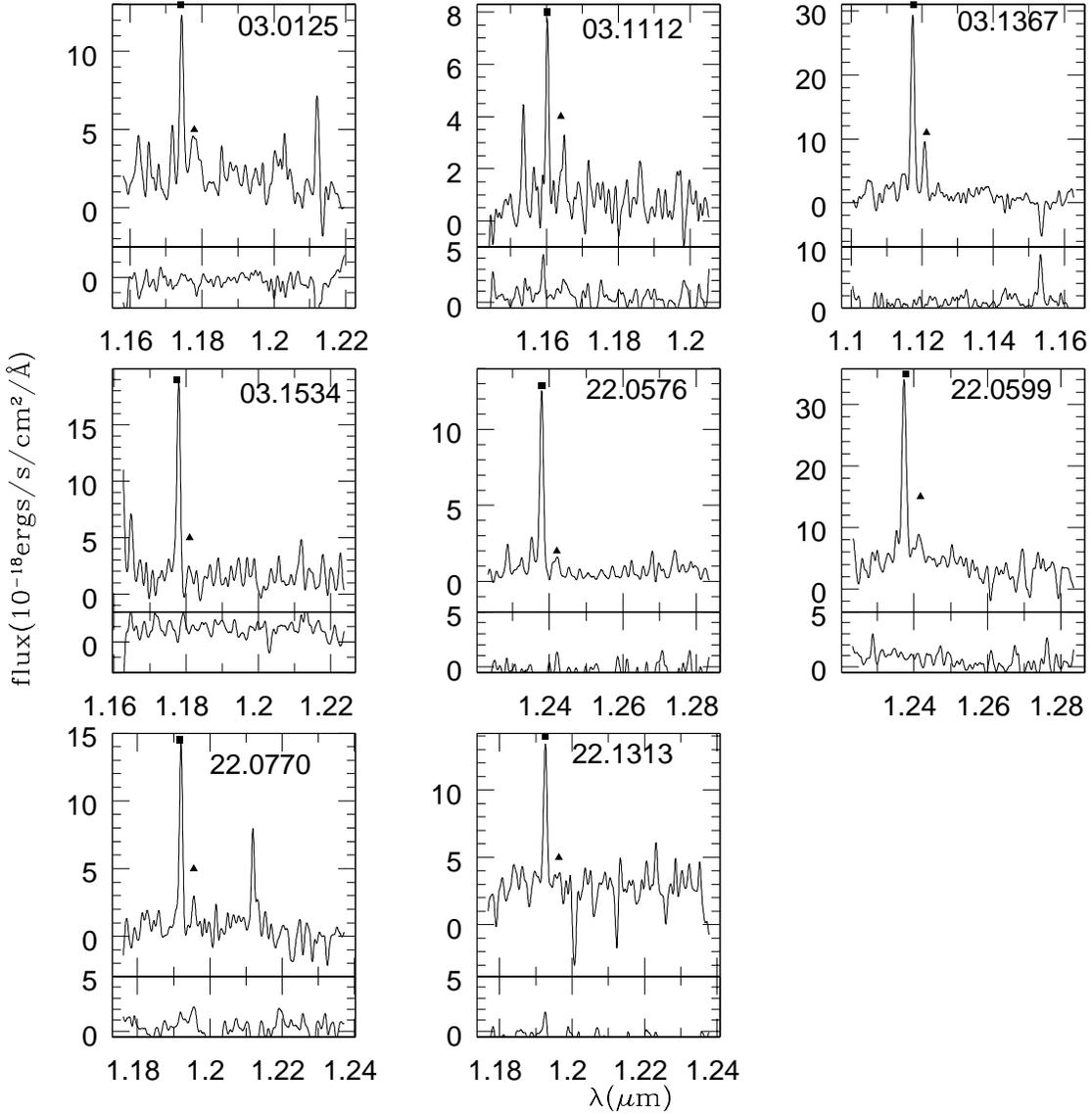}
\caption
{\label{specJ}\footnotesize Like Fig.\,\ref{specSZ}, but using the J
  filter for the
  4th grating order in order to get \Ha\, and  \NII\, for galaxies at
  $0.7<z<0.92$. Almost all galaxies show a \NII/\Ha\, ratio greater than 0.1, as
  expected for galaxies that lie on the high-metallicity branch of the
  \R23\, calibration.
}
\end{figure}

The one-dimensional spectra of each galaxy
were extracted 
 using the algorithm by \citet{horne} with an aperture of 10-15 pixels in y-direction, i.e. about
1.5$-$2\arcsec.  
An effective spectral resolution element of FWHM$\sim8$\AA\, in the J filter,
and FWHM$\sim5.9$\AA\, in the SZ filter, i.e., 13-14 pixels, was estimated
from the calibration lines.  We therefore smoothed the reduced 1D
spectra when displaying them in Figs.~\ref{specSZ} and \ref{specJ},  to
sample the spectral resolution element, thereby maximizing
the S/N ratio without any loss of information.
A Gaussian filter was used as a kernel for the smoothing, as a Gaussian shape
well reproduces the typical instrumental profile and thus minimizes any change
of shape in the profiles of the observed spectra lines.  

%************************

\subsection{The Keck---NIRSPEC Spectra}
Eleven galaxies were observed using NIRSPEC at Keck II on the nights of
June 14/15 2000,
 29/30 March 2004, and 30/31 March 2004.
NIRSPEC was designed primarily as a high-dispersion echelle spectrograph. We
used it in the low-dispersion mode, which replaces the echelle with a mirror,
so that the dispersion is provided by the order-sorting grating. The resulting
spectrum is tilted by about 6\arcdeg\ from the orientation of the 
$1024\times1024$ Alladin array columns, and the dispersion varies slightly
along the projected slit.

The objects were moved on the slit in an A-B-B-A pattern for each 4-exposure
sequence, each exposure being 600 s. For each exposure, the following
steps were taken before correcting for distortion, in order to optimize both
the airglow subtraction and the removal of cosmic rays. For, say, the first
exposure at the ``A'' position, the two exposures at the ``B'' position were
averaged, Gaussian smoothed slightly parallel to the airglow lines to minimize
added noise, scaled to match as closely as possible the airglow lines
in the region of redshifted H$\alpha$, and subtracted. We used the
{\it lacos\_im} task \citep{vDo01} to remove cosmic rays. We then added back
the subtracted spectrum. The resulting spectrum is simply the orginal spectrum
with cosmic rays removed.

After this procedure was carried out for each of the 4 spectra of a group, we
repeated the process, using the cosmic-ray-free spectra and retaining the
difference image, which was now free of cosmic rays (including the negative
cosmic-ray residuals from the original subtraction) and largely free of
airglow lines, although some slight positive and negative residuals remained.

At this point, we rectified the spectrum, corrected for distortion, and
applied the wavelength calibration (derived from the airglow lines), using
the {\it wmkonspec} IRAF package.
% developed by G.~Wirth. 
Airglow residuals 
were removed from the rectified spectra by fitting a cubic spline (with outlier
rejection) to each row (i.e., the cross-dispersion dimension) and subtracting.
The 1-D spectra were extracted from these images, some additional manual
cleaning of ``unfeasible noise'' was done, and the final spectra were 
averaged together.

\begin{figure}[h!]
\plotone{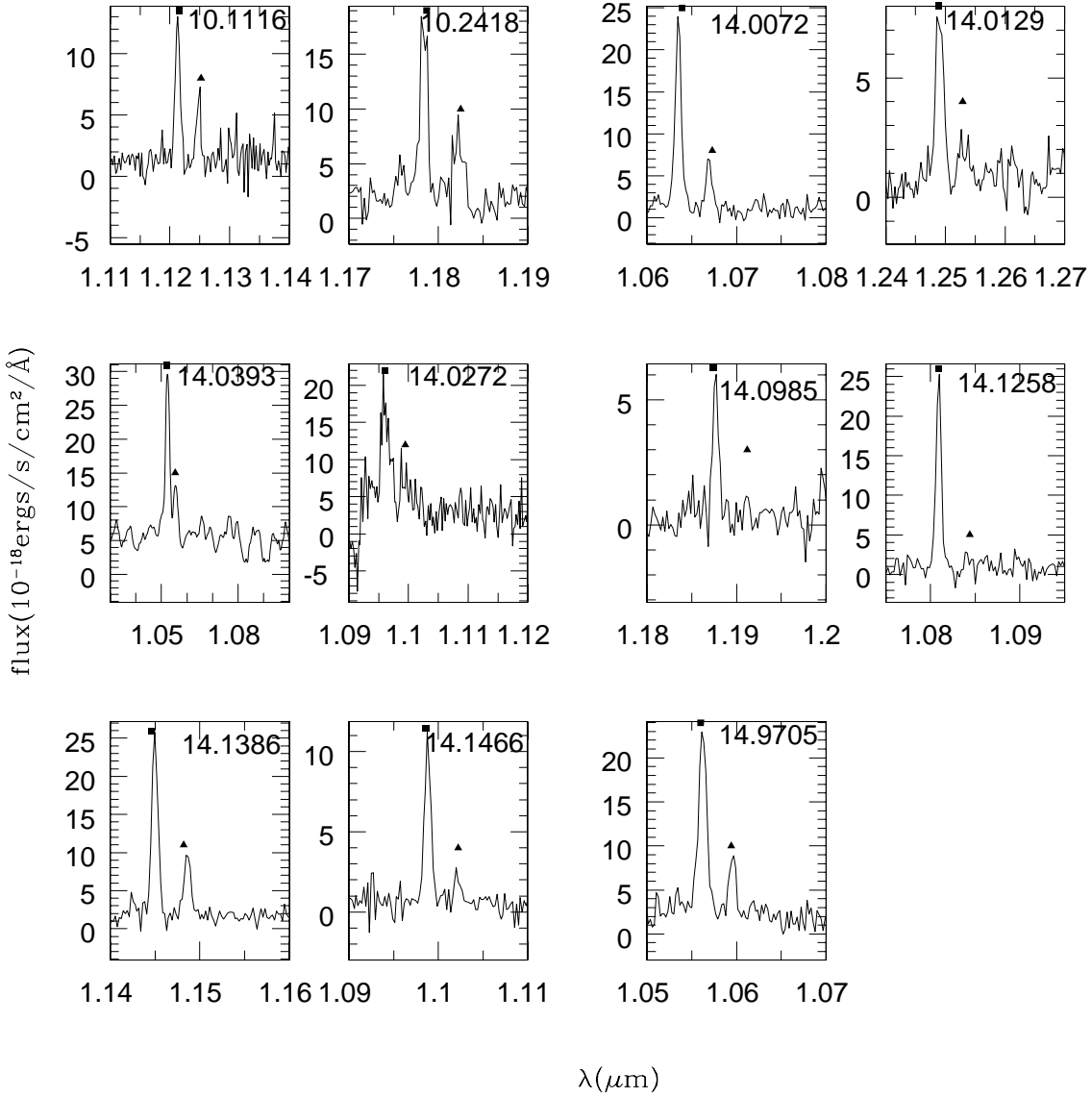}
\caption
{\label{specKeck} \footnotesize NIRSPEC  spectra using Keck. In each figure, the filled square and  the filled triangle show
  the position of \Ha\, and  \NII, respectively. Almost all galaxies show a \NII/\Ha\, ratio greater than 0.1, as
  expected for galaxies that lie on the high-metallicity branch of the
  \R23\, calibration.
}
\end{figure}

Final calibration and correction for telluric absorption was carried out with the
aid of observations of the A0 V stars HD\,89239 and 86 UMa, and the A2 star
HD\,129653. The procedures were similar to those described in Section \ref{speccal}, the main
difference being the need to interpolate across the Pa-$\beta$ or Pa-$\gamma$
lines in the standards in some cases.
The 1D  flux calibrated spectra from Keck are shown in Fig.\,\ref{specKeck}.

%************************

\subsection{Measurement of the \Ha\, and \NII\, fluxes, and line ratios}

The \Ha\, and \NII\, fluxes (see Table~\ref{infoobj}) were measured
using the
package \emph{splot} in IRAF.  The flux errors were usually dominated by
systematic uncertainties in establishing the local continuum. The latter was
conservatively estimated by exploring rather extreme possibilities.
As in LCS03, we assumed that the flux of \OIIIb\, is 0.34 of the flux of
\OIIIa. Similarly, we assumed an average underlying stellar absorption in
\Hb\, of $3\pm2$\AA\, and corrected the equivalent width of this line
by this, thereby increasing the \Hb\, line flux.
 The treatment of Balmer
  absorption propagates through to $A_{V}$, e.g., correcting too much
  for the \Hb\, absorption would lead to an underestimate of the
  reddening and vice versa. Using
  $3\pm2$\AA\, to correct the \Hb\, absorption, the error bars should
  reflect this uncertainty.
We have not applied any
  corection to the \Ha\, line which  has an observed equivalent width
larger than 50\AA\, for all 30 CFRS galaxies (see Table~\ref{infoobj}).

A key issue for the measurement of extinction 
is the  matching between the near-infrared and optical spectra.
We used comparable slits for  the optical and  the  near-infrared
spectroscopy, and in the good seeing conditions encountered
the resulting emission line fluxes
from the spectroscopy should approximate
``total'' fluxes.
To check the accuracy of the relative calibration between
H$\alpha$ measurements and the H$\beta$ fluxes measured with CFHT, we compared,
for the seven objects with a detectable continuum, the value of the continuum at the
position of H$\alpha$ with the continuum at H$\beta$. These two measurements
of the continuum were  found to be consistent at the 10\% level with the broad band
magnitudes of the target galaxies.

%#######################################################################################

\section{Our method to determine gas oxygen abundances}

%###############################################################################

 Our approach  to determine gas oxygen abundances is based on the models of KD02, who developed a
set of ionisation parameter and oxygen abundance diagnostics based  on the
use of only strong optical emission lines.  The method consists in performing a
simultaneous fit to the \OII, \Hb, \OIIIa, \Ha, and \NII\, lines in terms of
extinction parameter $A_{V}$, ionisation parameter $q$, and [O/H].

For each set of metallicities, KD02 give third order polynomial fits to the
relation between \OIIIa/\OII\, and the ionisation parameter $q$. These have
the form:

\begin{equation}
\rm{log}([OIII]/[OII])=k_{0} + k_{1} \rm{log}\,q + k_{2} (\rm{log}\,q)^{2} + k_{3} (\rm{log}\,q)^{3}  \label{eqOIIIOII}
\end{equation}
where  $k_{0-3}$ are constants given in Table~2 of KD02.

For each ionisation parameter, fourth order polynomials are used to fit 
the model curves relating the flux ratios (or \R23\,) and metallicity:

\begin{equation}
\label{poly4}
\log(R)=k'_{0} + k'_{1}x + k'_{2}x^{2} + k'_{3}x^{3} + k'_{4}x^{4}
\end{equation}
where $R$ is either [NII]/H$\alpha$, or R$_{23}$, $k'_{0-4}$ are constants
given in Table~3 of KD02, and the variable $x$ is the metallicity, 12+log\,(O/H).

Using these relations between \OIIIa/\OII, \NII/\Ha, and \R23, and the
relations between [O/H], $A_{V}$, and $q$, we created a model grid of
relative line strengths as a
function of three parameters: the extinction parameter, $A_{V}$, the
ionisation parameter, $q$, and the oxygen abundance, 12+log\,(O/H).  The
ranges for the values of the parameters, and the details of the adopted grid are
given in Table\,\ref{modelgrid}.  The grid contains a total of $\sim 12 \times
10^{6}$ models and is described in detail in the next section.

It should be noted that there might be  systematic uncertainties in the KD02 models,
but our philosophy is to treat all galaxies at high and low redshift  in the same way.
Starting with the
observed line fluxes (ratios) and computing oxygen abundances with the
same method for the low and high redshift galaxies  \citep[as also
suggested and as very important seen by][]{salzer05}   will  allow us 
to focus on relative effects between the selected
samples in the expectation that these are likely to be much more robust
than atempts to determine very accurate the absolute metallicity.

\subsection{Extinction and model line fluxes $M_{i}$} 

The dereddened value for a flux line ratio, $\rm{I}(\lambda 1)/\rm{I}(\lambda
2)$, is given by:

\begin{equation}
\frac{\rm{I}(\lambda 1)}{\rm{I}(\lambda 2)} = \frac{\rm{F}(\lambda
    1)}{\rm{F}(\lambda 2)} \times 10^{c(f(\lambda 1)- f(\lambda 2))},
\end{equation}  

\noindent where $\rm{F}(\lambda)$ is the observed flux at a given wavelength, c is the
logarithmic reddening parameter that describes the amount of reddening
relative to \Hb, and $f(\lambda)$ is the wavelength-dependent reddening
function \citep{whitford}. $f(\lambda)$ can be approximated in the whole
spectral range to an accuracy better than 5\% with the expression $f(\lambda)
= 3.15854 \times 10 ^ {-1.02109\lambda} - 1$, where $\lambda$ is in units of
$\mu$m, see \citet{izotov}.
The value of $c$ can be estimated from the relation $c = 1.47 \times
\rm{E}_{\rm{B-V}}= 1.47 \times A_{V}/3.2 $, and the dereddened flux of an
observed emission line is $F_{der}=F_{obs}\times 10^{c(1+f(\lambda))}$
\citep{seaton}.

The \Ha/\Hb\, line ratio is quite insensitive to the assumed density
and not very sensitive to the electron temperature, see Table\,4.4 in \citet{osterb}.
Assuming case B Balmer recombination, with a temperature of 10\,000\,K, and a
density of 100\,${\rm{cm}}^{-3}$ \citep{brockle}, the predicted dereddened
intensity ratio of \Ha\, to \Hb\, is 2.86  \citep{osterb}.  The effect of
reddening on the ratio \Ha/\Hb\, can be written as $\rm{I}(\Ha)/\rm{I}(\Hb) =
\rm{F}(\Ha)/\rm{F}(\Hb)\times 10^{(-0.332)c}$.

We normalized the model fluxes relative to \Ha\,, i.e., the \Ha\, flux was set
to one, and the correction for extinction assuming case B recombination
was calculated relative to \Ha.  
%The extinction corrected model value for \Hb\, is thus calculated according to:
The dust reddenened  model value for \Hb\, (to be compared with the
observed dust reddened emission line flux) is thus calculated according to:

\begin{equation}
\rm{F}(\Hb) =
\frac{\rm{F}(\Ha)}{2.86} \times 10^{-0.332\,c}.
\end{equation}

To get the model value for the \NII\, flux, \NII/\Ha\, is calculated
using equation \ref{poly4}, and then: 

\begin{equation}
\rm{F}(\NII) =
\frac{\rm{F}(\NII)}{\rm{F}(\Ha)} \times 10^{0.003\,c}.
\end{equation}  

To calculate the model value for the \OII\, flux, first \OIIIa/\OII\,
and \R23\, are calculated from equations  \ref{eqOIIIOII} and \ref{poly4}, respectively, and then:

\begin{equation}
\rm{F}(\OII) = 
\frac{\R23\, \times \rm{F}(\Hb)}{1.34 \times 10^{\rm{log}(\OIIIa/\OII)}+1} \times 10^{-0.640\,c}.
\end{equation}  

Using the calculated flux of \OII\, and the values for \R23\, from
equation  \ref{poly4}, we then obtained the model value for the
\OIIIa\, flux:

\begin{equation}
\rm{F}(\OIIIa) =
\frac{\R23 \times \rm{F}(\Hb) - \rm{F}(\OII)}  {1.34}  \times 10^{-0.298\,c}.
\end{equation}

Using the grid parameters given in Table\,\ref{modelgrid}, we thus
created about $ 12 \times
10^{6}$ models, each yielding 4 model flux ratios, or  equivalently  5
model fluxes $M_{i}$, $i=1,..5$, with  \Ha\, set to one ($M_{4}=1$) as described above.  

%**********************

\subsection{The best fit abundances, extinction values and ionisation parameters}
\label{bestfits}

For each of the 30 galaxies of our sample, combining the new \Ha\, and \NII\, measurements
with the available optical CFHT measurements of LCS03 provided us with five
emission lines, namely \OII, \Hb, \OIIIa, \Ha, and \NII. These are labelled
$D_{i}$, $i=1,..5$. For the comparison with the models, we assumed a
conservative minimum 10\% for the uncertainties  $E_{i}$ of the measured  fluxes, $D_{i}$.

The measured [OII], H$\beta$, [OIII], H$\alpha$, and [NII] fluxes (i.e.,
$D_{i}$, $i=1,..5$) were then compared with the theoretical fluxes $M_{i}$,
$i=1,..5$ predicted for each of the $\sim 12 \times 10^{6}$ models of our grid (which covers
a large range of $A_{V}$, $q$, and [O/H] values). In detail, 
the best fit models were selected to be those which minimized the 
$\chi^{2}$ function defined as:
\begin{equation}
\chi^{2} = \sum_{i=1}^{5} \left( \frac{c \cdot M_{i} - D_{i}}{E_{i}} \right)^{2},
\end{equation}  
where c is a flux normalization factor required to match the  model
fluxes $M_{i}$ and the measured fluxes $D_{i}$. The normalization $c$ is
chosen so that $\partial \chi^{2}/ \partial c=0$, which gives:

\begin{equation}
c= \left( \sum_{i=1}^{5}  \frac{D_{i} \cdot M_{i}}{E_{i}^{2}} \right) /
\left(\sum_{i=1}^{5}  \frac{M_{i}^{2}}{E_{i}^{2}} \right).
\end{equation}

For each of the 30 galaxies, the so-derived best fit models provide the oxygen
abundance [O/H], ionisation parameter $q$, and $A_{V}$.
The best fit [O/H] abundances  and  $A_{V}s$ which are used in the
analysis described in the following sections  are given in
Table\,\ref{CFRSHaNII2}.

\subsection{Error bars}
\label{erbars}
The error bars that we report in Table \ref{CFRSHaNII2}
are the formal 1$\sigma$ confidence intervals for the projected best-fitting values, yielding
oxygen abundances with a formal accuracy of typically $\la0.1$\,dex,
except for few objects, where the error bars are larger
  because these objects are near the turn-around region or because
  \NII\, is not able to break the \R23\,  degeneracy.
In detail, the error bars of the oxygen abundance (or $A_{V}$) are
given by  the range of oxygen abundance (or $A_{V}$, respectively) for those 
models with $ \chi^{2}$ in the range  $\chi_{min}^{2} \leq \chi^{2} \leq
\chi_{min}^{2} +1 $ (where  $\chi_{min}^{2}$  is the minimum $\chi^{2}$
of all allowed models for a given galaxy), corresponding to  a confidence level of 68.3\%
for one single parameter.

\begin{figure}[h!]
\plotone{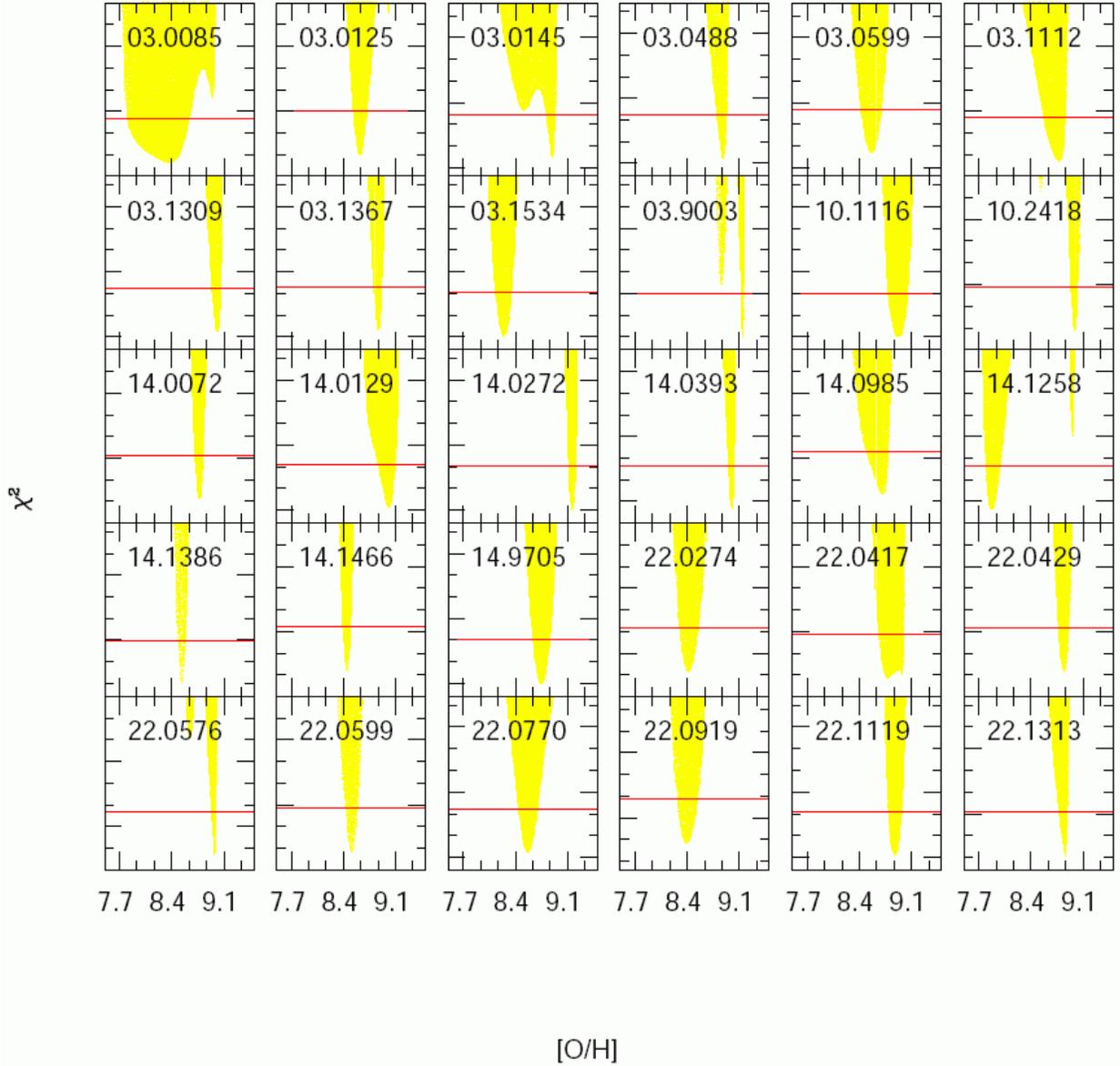}
\caption
{\label{chi2OH}\footnotesize $\chi^{2}$ values projected onto the [O/H]
  axis for the most probable  models  for each of the 30 CFRS galaxies.  The horizontal (red) line in
  each panel corresponds to $\chi_{min}^{2}+1$, where  $\chi_{min}^{2}$  is the minimum $\chi^{2}$
of all allowed models for a given galaxy. If only one peak (with a
  corresponding $\chi_{min}^{2}$) is
  intersected by this line, which is the case for almost all (28/30) CFRS galaxies,
  than the minimum and maximum oxygen abundances given by  
  the intersection points are  used to
  determine  the error bars of the oxygen abundance. 
}
\end{figure}

Figure~\ref{chi2OH} shows the $\chi^{2}$ values projected onto the
[O/H] axis, showing that generally only one metallicity solution is
possible, and also showing the formal range of allowed models. 
The horizontal (red) line in
  each panel corresponds to $\chi_{min}^{2}+1$.
 If only one peak (with a corresponding
  $\chi_{min}^{2}$) is
  intersected by this line, which is the case for almost all (28/30) CFRS galaxies,
  than the minimum and maximum oxygen abundances given by 
  the intersection points are  used to
  determine  the error bars of the oxygen abundance.

It should be noted that we consider a
  possible second solution corresponding to the
  second (higher)  peak,  only if the minimum $\chi^{2}$ of the higher peak
  lies in the range  $\chi_{min}^{2} \leq \chi^{2} \leq \chi_{min}^{2}
  +1 $.
There are two of the 30 CFRS galaxies where an alternate model is not strongly
  excluded ($ 1 <  (\chi^{2} - \chi_{min}^{2}) < 2 $). 
 In these two cases we also consider the alternative
  oxygen abundance solutions (corresponding to the higher, less
  probable peak)  and will indicate them as  open squares wherever
  [O/H] is plotted. The two
  alternative solutions are [O/H]$=8.51^{+0.19}_{-0.19}$ for object
  03.0145, and  [O/H]$=8.85^{+0.05}_{-0.04}$ for object 03.9003.

It should be also noted that the shape of the log\R23\, vs. [O/H]
  relation is flatter  in the high metallicity region (of the high
  metallicity branch) than in the turn-around and low metallicity region
  (see Figure 5 in KD02 - rotated by 90 degrees). Therefore the lower metallicity galaxies in our sample 
should have larger error bars than the high metallicity  galaxies, as
  we obtain.

In terms of absolute measurements of metallicity there are  three
source of uncertainties: 

(a) the purely statistical measurement
uncertainties propagating through to the parameter determinations. These
are addressed by our $\chi^{2}$ analysis.  They reflect both the quality 
of the data and the gradients (and degeneracies) in the models.
These uncertainties are presented in  Table \ref{CFRSHaNII2}. 

(b) Second, there are additional uncertainties arising from
the methodology - e.g., averaging together the light from all of the
HII regions in the galaxies to derive some kind of average [O/H]-
probably at the  $\pm 0.1$\,dex level \citep{kob99}.
 
(c) Uncertainties in the Kewley and Dopita models. Application 
of many different \R23\, calibrations to the SDSS data (Sara Ellison 
private communication, Ellison and Kewley 2005, submitted) indicates 
a range of $\pm 0.2$\,dex in the mean [O/H] at a given luminosity -
with the KD02 models more or less in the middle. This latter problem is 
reduced by using the same analysis on all objects and focussing on
differential effects with $z$ or L, rather than estimates of the absolute
metallicity.

Our philosophy is to treat all galaxies at high and low redshift  in
the same way and to focus on relative effects between the selected samples. 
Therefore, when comparing metallicities of the different
 samples,  we think it is more appropriate to  
consider  the uncertainty (a) described above.

%**********************

\subsection{Comparison with the oxygen abundances of LCS03}
\label{OHcomp}

\begin{figure}[h!]
\plotone{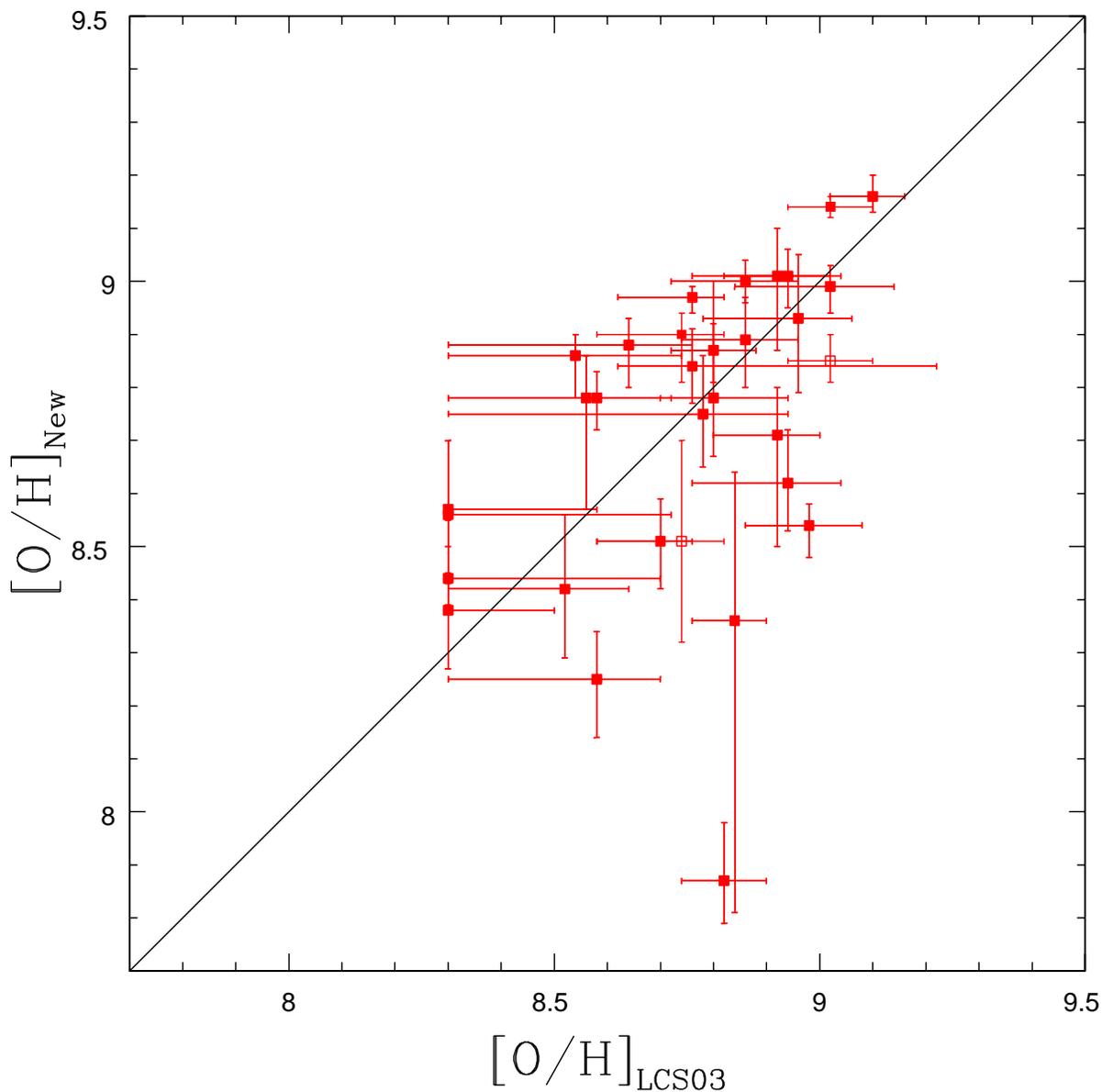}
\caption
{\label{OHcompfig} \footnotesize Comparison of the best fit [O/H]
  abundances derived with the new method for 30 CFRS galaxies (and based on the full set of
  five emission lines) with the [O/H] estimates derived from \R23\, and
  presented in LCS03. The two open squares are the alternative (but
  less probable) oxygen abundance solutions for two CFRS galaxies, as discussed in Section~\ref{erbars}.
In the  Sections \ref{OHcomp} and \ref{dustmet} we discuss  the physical basis and
the implications of the agreement and scatter seen in the figure.
}
\end{figure}

In Fig.\,\ref{OHcompfig} we compare the best fit [O/H] abundances derived with
the new method (and based on the full set of five emission lines) with the
[O/H] estimates derived from $R_{23}$ alone  assuming extinction $A_{V}=1$, and presented in LCS03.  Despite a
substantial scatter, Fig.\,\ref{OHcompfig} shows, on average, a good agreement
between the old and new [O/H] estimates. The discrepancies seen in   the figure between the LCS03 and
  our measurements based on the KD02 method
are due to the following: a) low \NII/\Ha\, ratios for 5/30 galaxies imply that  the upper
  branch assumption of LCS03 was incorrect for those 5 galaxies; 
 and b)  the extinction derived using the best fits of five emission
 lines are higher or lower  compared to the 
  $A_{V}=1$ assumed by LCS03,  affecting \R23\, and thus the derived
  oxygen abundance. We discuss the physical basis and
the implications of the agreement (and scatter) in  Fig.\,\ref{OHcompfig} in more detail in the discussion below.

%  or high  extinction and upper branch

%#######################################################################################
%
                             %RESULTS
%
%#######################################################################################
%\clearpage

\section{Results and Discussion: Metallicities, SFRs  and dust in normal galaxies at $z\sim1$}

\subsection{The nearby comparison sample}

As in LCS03, we choose the \citet[][NFGS]{jansen} galaxy sample as a local
comparison sample.  This sample was selected from the first CfA redshift
catalog,  and includes $z<0.04$ galaxies of all morphological
types. The galaxies  span 8 mag in luminosity and a broad range of environments. The
spectra are integrated over most of the luminous parts of the galaxies and
should thus be similar to the unresolved spectra of CFRS medium-redshift
galaxies.  From the  NFGS sample of 196 objects we have extracted the 108 galaxies 
with measured emission fluxes for the five lines of [OII], H$\beta$, [OIII],
H$\alpha$, and [NII].

Additionally we also use as a local comparison sample 70 galaxies from the KPNO
International Spectroscopic Survey \citep[KISS,][]{melbsal} which also have
measured line fluxes for the five emission lines mentioned above. KISS identifies
emission line galaxies candidates out to a redshift of $z=0.095$ by selecting
objects with detectable \Ha\, or \OIIIa\, in the low-dispersion
objective-prism spectra of the survey.  It should be noted that for the 70 KISS galaxies
only line \emph{ratios} are available: \OII/\Hb\, \citep[from][]{melbsal},
\OIII/\Hb, \NII/\Ha, and \Ha/\Hb\, (kindly provided to us by James Melbourne).
Therefore, for these 70 KISS galaxies we can calculate [O/H] using our method,
but not the (dust-corrected) star formation rates from the \Ha\, fluxes.

%**************************************************************

\subsection{Star formation or AGNs?}

\begin{figure}[h!]
\plotone{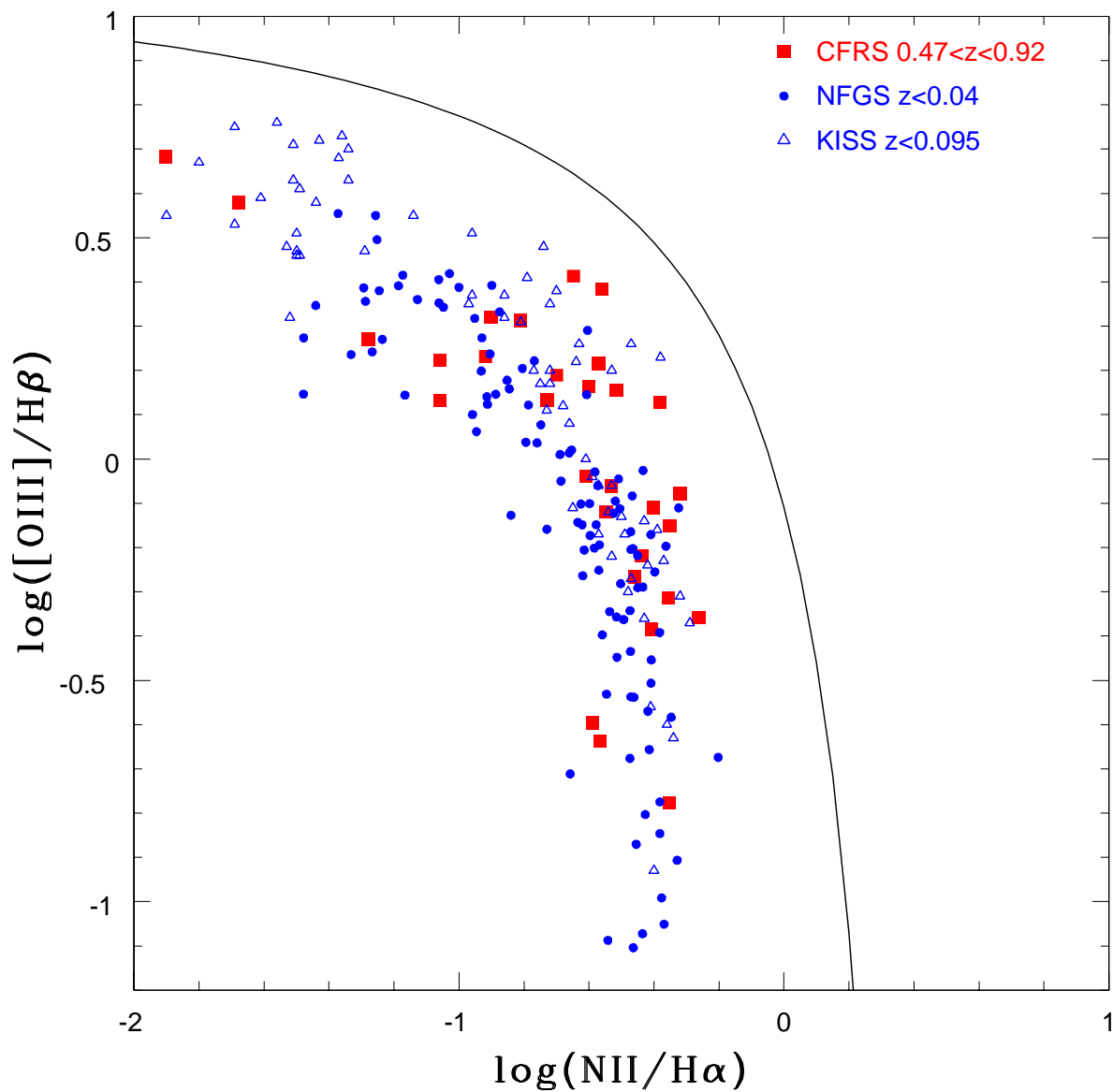}
\caption
{\label{linediag} \footnotesize Diagnostic diagram for
  the 30 CFRS galaxies at $0.47<z<0.92$ (filled squares), 108 local NFGS
  galaxies (filled circles), and 70 KISS
  galaxies (open triangles). 
The solid line shows the theoretical curve of \citet{kewley01} above
  which galaxies are dominated by an AGN. 
Similarly to the nearby star forming galaxies of the
combined NFGS+KISS comparison sample, all of the CFRS galaxies lie below the
theoretical curve, indicating that in all of them the dominant source of
ionisation in the gas is recent star formation.
}
\end{figure}

An essential step required for proceeding with the interpretation of the line
emission properties of the intermediate-z galaxies is to establish whether the
source of gas ionisation is of stellar origin, or rather associated with AGN
activity.  In order to identify galaxies possibly dominated by an AGN,
we can now use
the log(\OIIIa/\Hb) versus log(\NII/\Ha) diagnostic diagram presented in
Fig.\,\ref{linediag}.

AGN-dominated galaxies should appear above the solid line which represents the
theoretical threshold between star forming galaxies and AGNs predicted by
\citet{kewley01}.  Similarly to the nearby star forming galaxies of the
combined NFGS+KISS comparison sample, it should be noted that all of the CFRS galaxies lie below the
theoretical curve, indicating that in all of them the dominant source of
ionisation in the gas is recent star formation.

%****************************************************

%\clearpage
\subsection{\NII/\Ha\, as surrogate for oxygen abundance?}
\label{niihaoxyg}

\begin{figure}[ht!]
\plotone{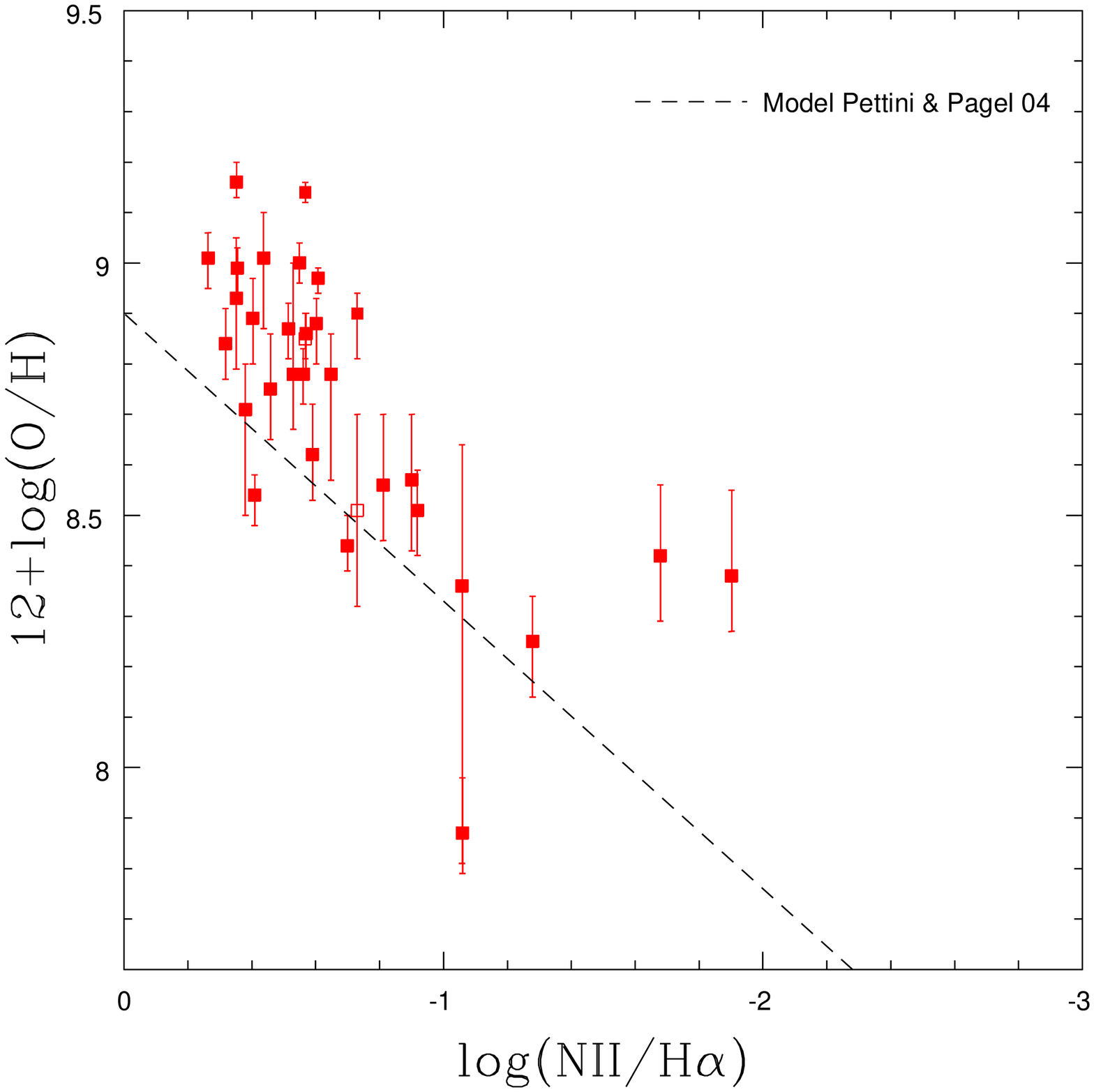}
\caption
{\label{NIIHa_OH} \footnotesize [O/H] based on 5 measured emission
  lines versus \NII/\Ha\,  for the  30 CFRS
  galaxies compared 
to the \citet{petpag} relation between \NII/\Ha\, and metallicity
(dotted line). 
As discussed in Section \ref{niihaoxyg}, \NII/\Ha\, can be used to determine the
upper/lower branch of the \R23\, relation, but it is only a very crude estimate of
the oxygen abundance.
}
\end{figure}

\citet{petpag}  derived an empirical calibration for
the oxygen abundance based only on the \NII/\Ha\, ratio:
$12+\rm{log(O/H)}=8.90+0.57 \times \rm{log}(\NII/\Ha)$.
 Fig.\,\ref{NIIHa_OH}
shows the oxygen abundances for the 30 CFRS galaxies with 5 emission
line fluxes measured versus their \NII/\Ha\, ratios, compared with the
\citet{petpag} relation between oxygen abundance and  the \NII/\Ha\, ratio.
It is obvious from this diagram that the \citet{petpag} relation is
only a very rough estimate of oxygen abundances.
This is consistent with the finding of KD02 based on  their Fig.\,7:
for log(\NII/\Ha)$>-0.8$ the [NII]/\Ha\, metallicity relationship
breaks down. Although such high [NII]/\Ha\, ratios indicate an oxygen
abundance on the upper branch of the \R23\, relation, \NII/\Ha\,
cannot be used to estimate a  metallicity in this regime. At lower
[NII]/\Ha\, values, the [NII]/\Ha\, ratio is less sensitive to
metallicity and more dependent on \R23\, and ionisation parameter. 
Therefore, \NII/\Ha\, can be used to determine the
upper/lower branch of the \R23\, relation, but it is only a very crude estimate of
the oxygen abundance.

%****************************************************

\subsection{Dust extinction, metal enrichment, and extinction-corrected SFRs}
\label{dustmet}

The good agreement seen in Fig.\,\ref{OHcompfig} between the [O/H] estimates based on $R_{23}$ and
the new more accurate estimates derived by including the \Ha\, and \NII\, fluxes into the
derivation of the gas abundances comes from two things.

First, the new NIR spectra allow us to determine the \NII\,/\Ha\, ratios for
the individual galaxies. These are expected to be $>0.1$ for galaxies that lie
on the high-metallicity branch of the $R_{23}$ calibration. Indeed, we find
such high values of \NII\,/\Ha\, for most of our CFRS galaxies, confirming the
assumption made in LCS03. Note however that 5/30 CFRS galaxies are on
the lower branch, but near the turn around region of the \R23\, relation.

Second, the best fits to the five emission lines discussed in Section \ref{bestfits}
provide a mean extinction $A_{V} \sim 1$ for the CFRS galaxies. This is
the same as the $A_{V}=1$ value that was assumed in LCS03 for
the whole sample, establishing that most of the metal-enriched $z\sim1$ galaxies
are indeed not heavily obscured by dust.

\begin{figure}[hb!]
\plotone{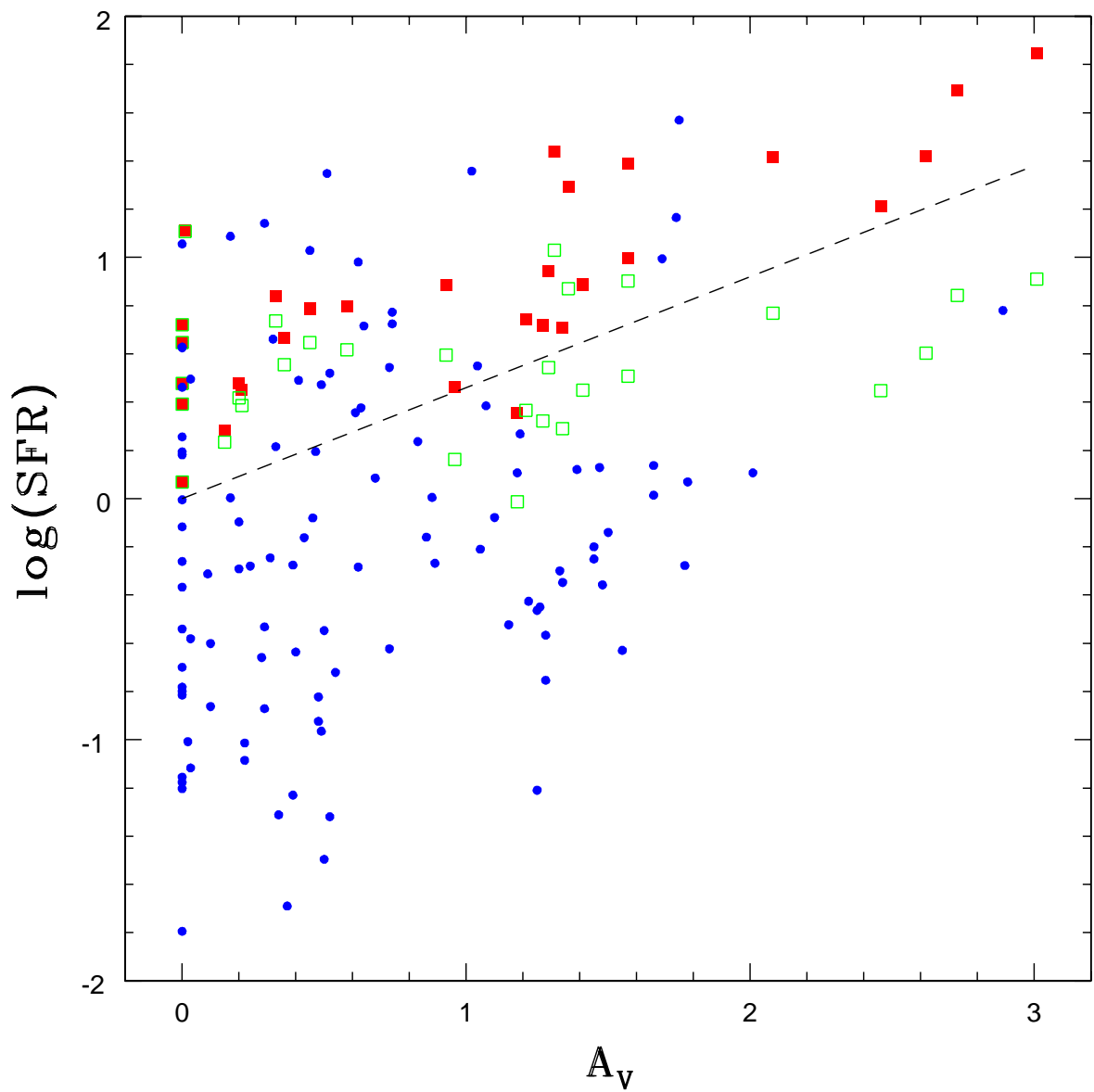}
\caption
{\label{AVSFR} \footnotesize Extinction corrected log(SFR) vs. A$_{\rm{V}}$ for the 30 
  CFRS (filled squares) and NFGS galaxies (filled circles).
The star formation rates \emph{not} corrected for extinction for the 30
  CFRS galaxies  (open squares)  show no correlation with A$_{\rm{V}}$. 
Therefore, if galaxies are lacking  measurements of \emph{both}  \Ha\, and \Hb\,
  to determine the extinction, one \emph{cannot} use   a similar A$_{\rm{V}}$
  value for galaxies with similar \emph{not} extinction-corrected \Ha\, luminosities (SFRs). 
  It is  important to determine the extinction for every single galaxy
  in order to obtain reliable measurements for the SFRs.
}
\end{figure}

On the other hand, the scatter around the mean $A_{V}$ value that we
derive from the NIR data is large, as shown in
Fig.\,\ref{AVSFR}. In detail, the assumed $A_{V}=1 \pm 0.5$ in
LCS03 is correct for only 11 of the 30 CFRS galaxies ($\sim37$\%), is
overestimated for 12 galaxies ($\sim40$\%), and is underestimated for 7 ($\sim23$\%
of the sample).
  This indicates that it is important to determine the
extinction for every single galaxy in order to obtain reliable
gas oxygen abundances in high redshift galaxies.

%**************

The availability of the individual extinction values $A_V$ for our sample
galaxies allows us to derive extinction-corrected SFRs from the \Ha\, fluxes.
For each CFRS galaxy, we calculated the SFR using the \citet{ken98} conversion
of H$\alpha$ luminosity into $\dot{\rm{M}}$: $\rm{SFR} (M_{\odot}\rm{yr}^{-1})
= 7.9 \times 10^{-42} \rm{L}(\rm{H}\alpha)\rm{ergs/s}$.
The resulting extinction-corrected SFRs of the 30 CFRS galaxies  range between $1$ and
$70\,M_{\odot}\rm{yr}^{-1}$. 
Fig.\,\ref{AVSFR} shows the SFR vs. extinction for the 30 CFRS galaxies
(filled squares)
and local NFGS galaxies (filled circles) indicating that 
CFRS galaxies have extinction properties broadly similar to nearby galaxies
with similar SFRs and luminosities.

We did not find strong
correlations between any pairs of  galaxy properties, from  absolute $B$ magnitude
$M_{B,AB}$, oxygen abundance [O/H], restframe $(U-V)_{AB,0}$  color, extinction
$A_{V}$, extinction-corrected SFR, and ionisation parameter $q$. This was
established by performing a Principal Component Analysis using these
quantities as Eigenvectors. It should be noted, however,   that the range of measured
parameters within the sample and the number of galaxies are both quite small.

The only strong  correlation found by this analysis was between extinction $A_{V}$ and the extinction-corrected
SFR, as seen also in  Fig.\,\ref{AVSFR}.  Obviously, such a relation
could have been produced by an error in $A_{V}$: if the $A_{V}$
estimates were randomized, given the narrow range in \Hb\,
fluxes in our sample, we would still get a distribution with a mean relation parallel to the direction of the error vector
(dotted line in  Fig.\,\ref{AVSFR}).
We do not think that this is the case.

The CFRS sample was selected by LCS03 to have an equivalent width
$EW(\Hb)>8$\AA, and the sample of our 30 CFRS galaxies has a  selection
against objects below the dotted line. However, there is no selection
against galaxies with low extinction and high star formation rates,
which would appear in the upper left corner of  Fig.\,\ref{AVSFR}, but
which are apparently missing.

The star formation rates \emph{not} corrected for extinction
for the 30   CFRS galaxies  (open squares in Fig.\,\ref{AVSFR}, derived from \Ha\, luminosities not corrected for extinction) show no correlation with A$_{\rm{V}}$. 
Therefore, if galaxies are lacking  measurements of the Balmer lines
  required to determine the extinction, assuming  a similar
  A$_{\rm{V}}$ for galaxies with similar (\emph{not}
  extinction-corrected) \Ha\, luminosities  would lead to wrong estimates of the real SFRs. 
  This suggests that it is important to determine the extinction for every single galaxy
  in order to obtain reliable measurements for the SFRs of high
  redshift galaxies.

%#######################################################################################

\subsection{The metallicity-luminosity (mass) relation}

\begin{figure}[h!]
\plotone{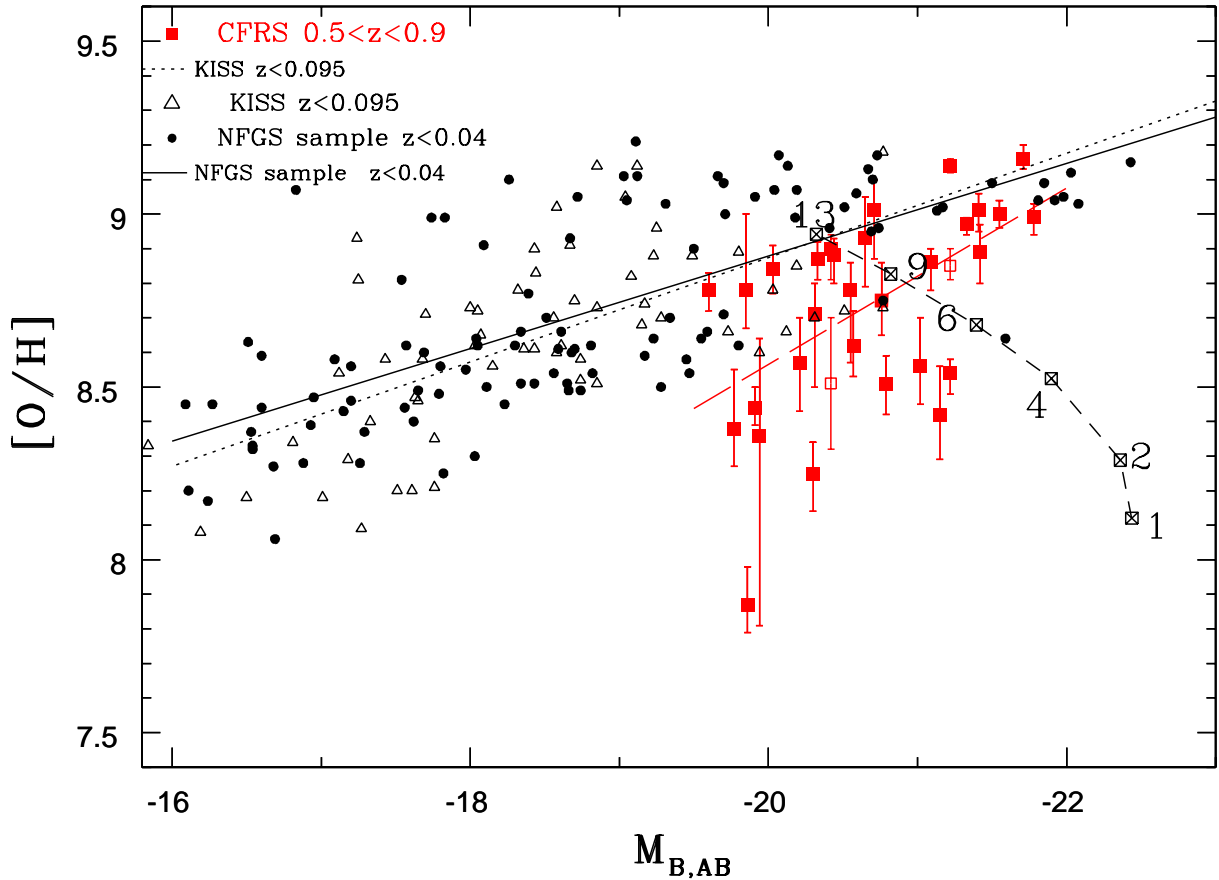}
\caption
{\label{MB_OH} \footnotesize Oxygen abundance versus $M_{B,AB}$ for
  the 30 CFRS galaxies (filled squares),  NFGS local galaxies (filled
  circles), and local KISS galaxies (open triangles). The dashed line shows the resulting
  metallicity-luminosity relation  of the 30 CFRS galaxies. The two open squares are the alternative (but
  less probable) oxygen abundance solutions for two CFRS galaxies, as discussed in Section~\ref{erbars}.
The lower metallicity CFRS
  galaxies are consistent with evolving into NFGS and KISS galaxies
  with slightly (0.5-0.9\,mag) lower luminosities and higher
  metallicities (a factor of  $\sim2$), as shown by an example scenario
 discussed by \citet{maier04}; note that the symbols along the track indicate the
  age of the model galaxy.
}
\end{figure}

\begin{figure}[h!]
\plotone{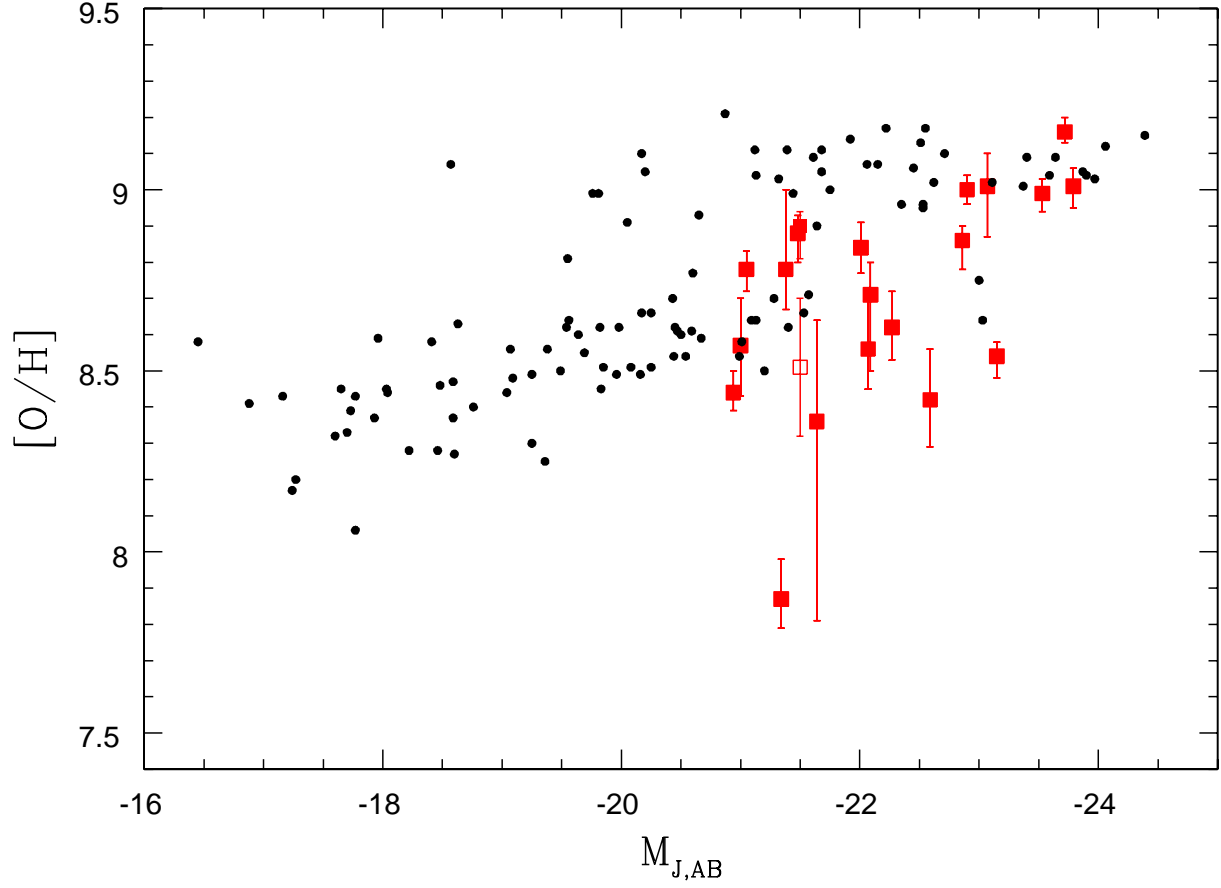}
\caption
{\label{MJ_OH} \footnotesize Oxygen abundance versus $M_{J,AB}$ for
  the 20 CFRS galaxies with available K photometry (filled squares), and NFGS local galaxies (filled
  circles). The lower metallicity CFRS
  galaxies are consistent with evolving into NFGS and KISS galaxies
  with slightly  lower luminosities and higher
  metallicities (see also Fig.\,\ref{MB_OH}).
}
\end{figure}

A metallicity-luminosity relation is observed in the local universe
\citep[e.g.,][]{melbsal,lamar04,tremon04}, in the sense that more luminous galaxies
tend to be more metal-rich.  Fig.\,\ref{MB_OH}, an update of Fig.\,10
in LCS03, shows  $M_{B,AB}$ vs. [O/H]
for the local NFGS and KISS galaxies and the fits to the respective data,
which result in a metallicity-luminosity relation of similar slope and
zero-point for both samples, which we use as local comparison.
The 30 CFRS galaxies with measured \Ha\, are plotted as filled squares in this
diagram. 
Moreover, as an update of Fig.\,17 in LCS03, Fig.\,\ref{MJ_OH} shows $M_{J,AB}$ (a surrogate for the
stellar mass) vs. [O/H] for the 20
galaxies with available K photometry.

\begin{figure}[h!]
\plotone{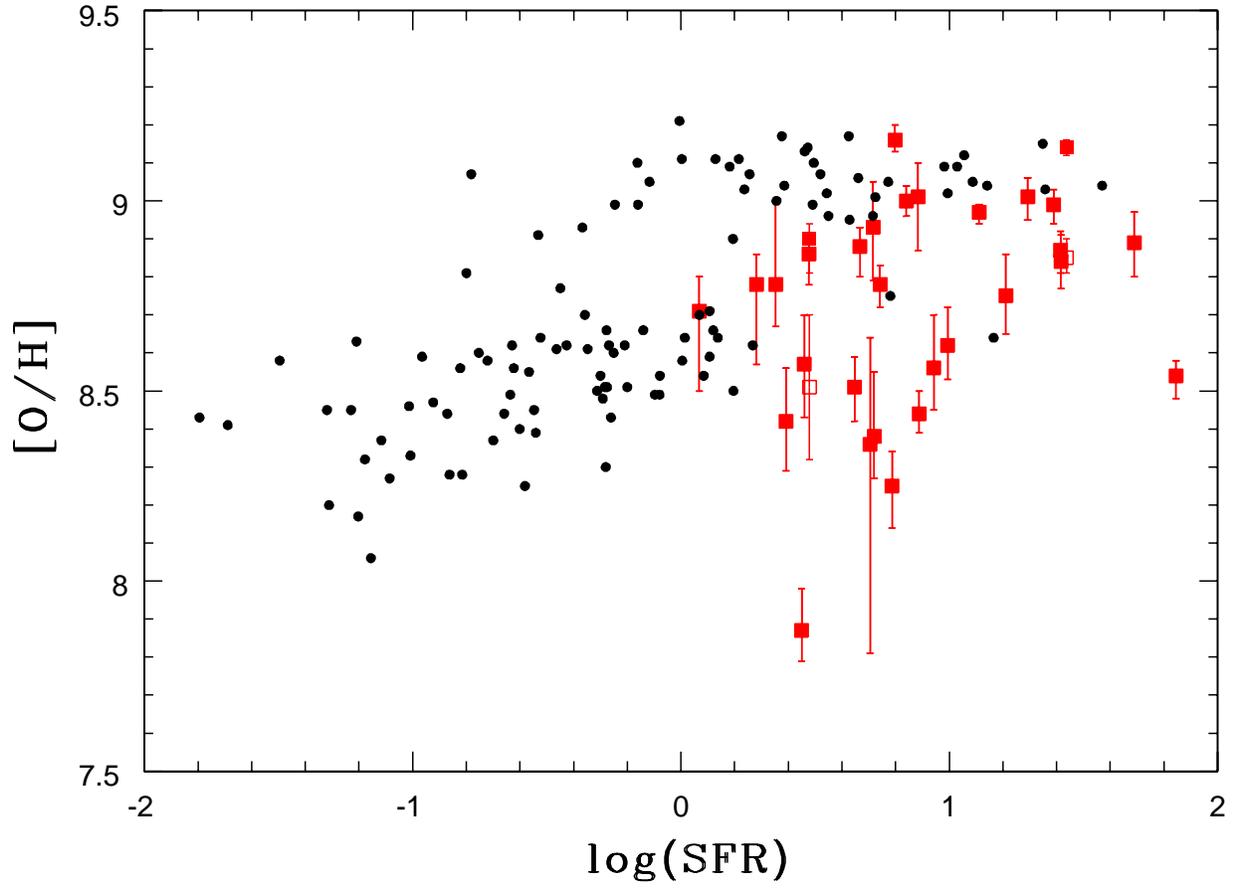}
\caption
{\label{OHSFRJan}\footnotesize Oxygen abundance [O/H] vs. log(SFR) for
  the 30 CFRS galaxies (filled squares), and   for  the local  NFGS
  galaxies (filled circles). The lower metallicity CFRS
  galaxies are consistent with evolving into NFGS and KISS galaxies
  with slightly  lower luminosities and higher
  metallicities, due to decreasing levels of star formation (see also Fig.\,\ref{MB_OH}).
}
\end{figure}

In both diagrams the higher
metallicity CFRS galaxies overlap with the region of the diagram
occupied by local galaxies of similar luminosity (mass). The lower
metallicity CFRS galaxies ($[O/H]< 8.6$) are more
luminous (massive) than local galaxies with similar [O/H], and more metal-poor than local
galaxies with similar absolute luminosities.
A similar trend is seen in Fig.\,\ref{OHSFRJan} for the SFR-[O/H]
relation: the  CFRS galaxies with higher metallicities overlap with  the region of the diagram
occupied by local galaxies with similar SFRs, while the lower
metallicity CFRS galaxies ($[O/H]< 8.6$) have higher SFRs than
local galaxies with similar [O/H], and are more metal-poor than local
galaxies with similar SFRs.

Comparing the average value of [O/H]
   for the 30 CFRS galaxies to the average value of [O/H] of
  NFGS and KISS local galaxies with similar luminosities in Fig.\,\ref{MB_OH}, we find that  the average change in
  metallicity is  about 0.3\,dex between galaxies at  $z\sim0.7$  and
  $z\sim0$. It is not hard to get a change by a factor of two between
   $z\sim0.7$ and $z\sim0$. For instance,  P\'egase2 models
   \citep{fiocrocca99} discussed by \citet{maier04} can get a  track as
   shown by the dashed curve and symbols in Fig.\,\ref{MB_OH},
   acording to which  the metallicity of the intermediate redshift CFRS galaxies
   may increase by a factor of about 2 by $z\sim0$. 
In this scenario the lower metallicity CFRS
  galaxies may fade by 0.5-0.9\,mag by $z\sim0$, due to decreasing
  levels of star formation, and migrate in the metallicity-luminosity
  and metallicity-SFR diagram
  (Figs.\,\ref{MB_OH}, \ref{MJ_OH}, and \ref{OHSFRJan}) to the region occupied by local galaxies with 
  lower luminosities and higher [O/H].

Assuming a uniform $A_{V}=1$ LCS03 found $\sim 25$\% of the galaxies in their
CFRS sample to have lower metallicities, i.e. [O/H]$< 8.6$, but pointed
out that if  $A_{V}=0$ for all galaxies in their sample the fraction of
galaxies with  [O/H]$<8.6$ dropped to 5\%. Our new data, which allow the
determination of extinction for each individual galaxy, confirm a high
average $A_{V}$, and, not surprisingly, a high fraction, actually about one
third (but in a smaller sample than those used by LCS03),  of lower metallicity galaxies (see  Fig.\,\ref{MB_OH}).       

\citet{kob03} apparently did not find the medium redshift $M_{B}\sim -21$ galaxies with lower
metallicities [O/H]$<8.6$ in their sample.
This could be explained by the fact that \citet{kob03} did no
 corrected  the fluxes of their  $0.26<z<0.82$ galaxy sample for
 extinction. Correcting for reddening would increase \R23\,   and move
 objects with apparently high metallicity to lower metallicity. Also, if not all
 the objects were on the upper branch, as assumed by \citet{kob03} based on \NII/\Ha\, measured 
  for  9 ($z<0.4$) galaxies of their 64 objects, might  result in a lower   metallicity for
  some of the galaxies.

Our new measurements of \Ha\, and \NII\, confirm the low metallicity 
population of galaxies at $M_{B}\sim-21$ found by LCS03. It is unlikely
\citep[see also LCS03, and][]{maier04} that these  low metallicity galaxies
are the progenitors of today's metal-poor dwarf galaxies  ($M_{B}\sim
-17$), because they would need to fade too much
compared with the observational results of the evolution of
the luminosities of galaxies at $0<z<1.5$: e.g.,  \citet{wolf03} found a
maximum fading of  2\,mag  between $z\sim 1.2$ and $z\sim0$.
Moreover, chemical evolution models 
generally produce rather ``oblique'' than ``horizontal'' tracks:   
the evolution in the metallicity-luminosity
diagram would be from bottom right to upper left, with galaxies evolving from low metallicity and high luminosities
towards higher metallicities and fainter luminosities, as shown by the
model in Fig.\,\ref{MB_OH}.

  A more detailed discussion of the
  metallicity-luminosity relation at $0<z<3$ including our additional
  measurements of oxygen abundances of galaxies at $z\sim1.4$ and comparison to theoretical chemical
  evolution models will be given in a separate paper
  \citep[][Paper IV in our series]{maier05}.

%#######################################################################################

\section{Conclusions}

%#######################################################################################

The metallicity of the star forming gas has been measured for 30 CFRS
galaxies with $0.47<z<0.92$ using optical CFHT and near-infrared VLT-ISAAC and
Keck-NIRSPEC spectroscopy.
Using the measurements of five emission lines it was possible to
determine the extinction, oxygen abundances and extinction corrected
star forming rates for these 30 luminous ($M_{B,AB}\la-19.5$) galaxies.
 The sample   was extracted from the  66 CFRS galaxies for which  LCS03
 obtained estimates of oxygen abundances based on 3 emission line
 fluxes (\OII, \Hb, and \OIIIa).
The additional \Ha\, and \NII\, obtained by the near-infrared
spectroscopy lead us to   the following conclusions:

~~~ 1. The source of gas ionisation in the 30 CFRS galaxies is not
associated with AGN activity, as derived from the log(\OIIIa/\Hb)
versus log(\NII/\Ha) diagnostic diagram. 

~~~ 2. The mean extinction $A_{V}\sim 1 $ for the 30 CFRS galaxies is
the same as the uniform $A_{V}=1$ value assumed by LCS03.  However,  the
large scatter $0<A_{V}<3$  indicates the importance  of the determination of  the extinction for
every single galaxy, to obtain   reliable oxygen
abundances and star forming rates at high redshift.

~~~ 3. Most galaxies have $\NII/\Ha>0.1$ confirming the assumption made
in LCS03 that most CFRS galaxies lie on the high-metallicity branch of
the \R23\, calibration. However, a minority of them (5/30) lie on the lower
branch, although near the turn-around region.

~~~ 4. 20 of the 30  CFRS galaxies at  $0.47<z<0.92$ have the higher
metallicities ([O/H]$>8.6$) found
locally in galaxies of similar luminosities. However, one third of the
CFRS galaxies have substantially lower metallicities
than local galaxies with similar luminosities and star formation rates.
This is at the upper band of the range  found by LCS03 for the fraction
of lower metallicities objects, and is due to the fact that we can
account for  the variety of reddening when computing the oxygen abundances. 
We also find that  the average change in
  metallicity is  about 0.3\,dex between the CFRS galaxies  and   local
  galaxies of similar luminosities.

~~~ 5. The evolution of the lower metallicities CFRS galaxies will be
probably  oblique in the metallicity-luminosity diagram: these galaxies
will probably increase their metallicities by about 0.3\,dex and decrease their
luminosities by about 0.5-0.9\,mag, evolving into the region occupied
by today's $z\sim0$ galaxies. Therefore they are unlikely be the progenitors of
the  metal-poor dwarf  galaxies   seen today.

%#######################################################################################

\acknowledgments
We would like to thank Lisa Kewley for pursuing with us the problem of
truncated coefficients in the KD02 paper, and for kindly sending us  the
coefficients with more decimal  places which accurately reproduce the
KD02 theoretical models.
We also want to thank James Melbourne for kindly sending us \NII/\Ha\,
and \Ha/\Hb\, for KISS galaxies. 
We are also grateful the anonymous referee for his/her suggestions which have
improved the paper. 
CM acknowledges support from the Swiss National Science Foundation.

%##########################################################################################

%##########################################################################################
%
%                               %FIGURES
%
%###########################################################################################
\clearpage

%##########################################################################################
%
                                %SFR-O/H Plot
%
%##########################################################################################
\clearpage

%###########################################################################################
\clearpage

%##########################################################################################

\clearpage

%###########################################################################################
\clearpage

%###########################################################################################
%
                                %AV - Ha, OH, SFR Plots
%
%###########################################################################################

%###########################################################################################
%\clearpage

%###########################################################################################
%
%                             MB - OH  -  Diagrams 
%
%###########################################################################################

\clearpage

%###########################################################################################

\clearpage

%##########################################################################################
%
                                %SFR-O/H Plot
%
%##########################################################################################
\clearpage

%##########################################################################################
%
%                               %TABLES
%
%###########################################################################################

%##########################################################################################
%
%                               %OBJECTS 
%
%###########################################################################################

\clearpage
\begin{deluxetable}{cccccccccccccccccl}
%\addtolength{\hoffset}{-2.5cm}
%\addtolength{\textwidth}{3cm}
%\addtolength{\voffset}{-1.5cm}
%\addtolength{\textwidth}{1.5cm}
\tabletypesize{\scriptsize \bf}
\setlength{\tabcolsep}{0.02in}
\tablewidth{0pt}
\rotate
%\tablenum{}
%\tablecolumns{}
\tablecaption{\label{infoobj}   The 30 $0.47<z<0.92$ CFRS galaxies
  }
\tablehead{
\colhead{Obj}   &    \colhead{redshift}    &    \colhead{U}  &  \colhead{B}    &   \colhead{V}
  &  \colhead{R}    &  \colhead{I}    & \colhead{Z}     &  \colhead{K}
  &  \colhead{log($r_{0.5}$)\tablenotemark{a}}&    \colhead{$M_{B,AB}$}&
  \colhead{$M_{J,AB}$} &       \colhead{\OII\tablenotemark{b,c}}                &     \colhead{\Hb\tablenotemark{b,c}}        & \colhead{\OIIIa\tablenotemark{b,c}}         &  \colhead{\Ha\tablenotemark{b}}         & \colhead{\NII\tablenotemark{b}}& \colhead{$EW_{\Ha}$(\AA)}
}
\startdata
% \tableline
03.0085  & 0.609 & 23.51 & 23.39 & 22.91 & 22.17 & 21.80 & 21.62 & 21.32 &   *   & -19.94 & -21.64  &  8.80$\pm$0.33 &4.57 $\pm$2.0    &  7.70$\pm$0.20 & 16  $\pm$2   &  1.4  $\pm$ 1.2 & 60 \\              
03.0125  & 0.789 & 23.72 & 23.45 & 23.19 & 22.48 & 21.84 & 21.67 & 20.87 &  0.61 & -20.57 & -22.27  & 10.10$\pm$0.26 &4.85 $\pm$1.4    &  1.30$\pm$1.2  & 14  $\pm$2   &  3.6  $\pm$ 0.8 & 80 \\              
03.0145  & 0.603 & 22.55 & 22.48 & 22.22 & 21.58 & 21.32 & 21.30 & 21.10 &  0.74 & -20.42 & -21.50  & 17.70$\pm$0.66 &6.21 $\pm$0.54   &  8.80$\pm$0.40 & 22  $\pm$3   &  4.1  $\pm$ 1.3 & 100 \\              
03.0488  & 0.605 & 22.87 & 22.69 & 22.61 & 21.90 & 21.64 & 21.57 & 21.31 &  0.32 & -20.44 & -21.48  & 25.50$\pm$0.66 &9.06 $\pm$1.03   &  13.80$\pm$0.33& 30$\pm$   3  &  7.5$\pm$ 1.5   & 60 \\  
03.0599  & 0.479 & 22.59 & 22.44 & 21.94 & 21.36 & 21.18 & 21.11 & 21.25 &  0.68 & -20.21 & -21.00  & 20.50$\pm$0.66 &4.17 $\pm$0.78   & 9.00$\pm$ 0.40 & 21.3$\pm$1.0 &  2.68 $\pm$ 2.0 & 130 \\        
03.1112  & 0.768 & 23.60 & 23.61 & 23.46 & 23.03 & 22.45 & 22.29 & *     &   *   & -20.55 &      *  &  8.50$\pm$0.33 &2.62 $\pm$1.1    &  7.00$\pm$0.33 & 8   $\pm$2   &  1.8  $\pm$ 0.7 & 100 \\              
03.1309  & 0.617 & 22.24 & 22.24 & 21.87 & 21.19 & 20.99 & 20.87 & 19.32 &  0.95 & -21.41 & -23.79  & 25.70$\pm$0.33 &13.00$\pm$2.00   &  5.90$\pm$0.40 & 59  $\pm$5   &  32.2 $\pm$ 3.2 & 70 \\              
03.1367  & 0.703 & 22.76 & 22.42 & 22.44 & 21.74 & 21.37 & 21.29 &     * &   *   & -20.33 &      *  & 10.50$\pm$0.40 &5.82 $\pm$0.42   &  9.20$\pm$0.40 & 34  $\pm$3   &  10.4 $\pm$ 0.7 & 150 \\              
03.1534  & 0.794 & 23.17 & 23.19 & 23.16 & 22.73 & 22.24 & 22.26 & *     &  *    & -20.30 &      *  & 17.00$\pm$0.33 &5.73 $\pm$0.35   &  10.80$\pm$0.26& 19$\pm$   2  &  1  $\pm$ 0.4   & 75 \\  
03.9003  & 0.618 & 22.77 & 22.41 & 22.00 & 21.20 & 20.83 & 20.63 &     * &  *    & -21.22 &      *  & 12.00$\pm$0.20 &10.10$\pm$1.78   &  2.60 $\pm$0.40& 85$\pm$   8  &  23 $\pm$ 3     & 70 \\    
\hline
10.1116  & 0.709 &     * &     * &     * &     * &     * &   *   & *     &    *  & -20.65 &      *  & 7.1  $\pm$0.30 &2.66 $\pm$1.0    & 2.00$\pm$1.5   & 11.9$\pm$1   &  5.3  $\pm$0.7  & 95 \\           
10.2418  & 0.796 &     * &     * &     * &     * &     * &   *   & 20.14 &    *  & -21.78 & -23.53  &7.05  $\pm$0.88 &5.53 $\pm$0.55   &3.00 $\pm$2.0   & 34  $\pm$3   &  15   $\pm$ 1   & 105 \\               
\hline
14.0072  & 0.621 & 23.73 & 23.63 & 23.32 & 22.76 & 22.42 & 22.45 & 21.30 &  *    & -19.60 & -21.05  &11.20 $\pm$0.26 &4.22 $\pm$0.22   &10.80$\pm$0.26  & 18.2$\pm$1.5 &  5.0  $\pm$ 0.7 & 160 \\        
14.0129  & 0.903 & 24.57 & 24.28 & 23.82 & 23.37 & 22.48 & 22.30 & 20.63 &  *    & -20.71 & -23.07  &5.00  $\pm$0.20 &3.04 $\pm$0.9    &2.0  $\pm$2.0   & 12.3$\pm$1.0 &  4.5  $\pm$ 0.5 & 200 \\        
14.0272  & 0.670 & 22.58 & 22.29 & 21.75 & 21.11 & 20.47 & 20.27 & 19.28 &  *    & -21.71 & -23.72  & 6.90 $\pm$0.26 &7.65 $\pm$1.17   & 1.60$\pm$0.40  & 27  $\pm$3   &  12   $\pm$ 2   & 60 \\               
14.0393  & 0.603 & 22.18 & 21.93 & 21.47 & 20.83 & 20.46 & 20.34 & 19.56 &  0.75 & -21.55 & -22.90  &    30$\pm$0.35 &13.38$\pm$1.52   & 10.19$\pm$0.57 & 46  $\pm$3   &  13   $\pm$ 4   & 80 \\              
14.0985  & 0.809 & 24.47 & 24.38 & 23.85 & 23.50 & 22.53 & 22.29 & 21.42 &  0.44 & -20.31 & -22.09  & 8    $\pm$1    &2.83 $\pm$0.7    & 3.80$\pm$0.40  & 4.8 $\pm$0.5 &  2    $\pm$ 1.0 & 105 \\   
14.1258  & 0.647 & 24.12 & 23.95 & 23.64 & 22.94 & 22.64 & 22.46 & 21.58 &  0.26 & -19.86 & -21.34  &9.2   $\pm$0.5  &5.63 $\pm$0.22   &7.7  $\pm$0.7   & 17.2$\pm$1.0 &  1.5  $\pm$ 1.0 & 190 \\      
14.1386  & 0.744 & 23.40 & 23.11 & 22.57 & 21.94 & 21.38 & 21.11 & 19.84 &  *    & -21.22 & -23.15  & 9.32 $\pm$0.18 &6.07 $\pm$0.78   & 2.79$\pm$0.29  & 41  $\pm$4   &  16   $\pm$ 2   & 100 \\               
14.1466  & 0.674 & 23.87 & 23.75 & 23.34 & 22.68 & 22.38 & 22.24 & 21.33 &  0.45 & -19.91 & -20.94  &11.80 $\pm$0.66 &3.21 $\pm$0.27   & 5.20$\pm$0.26  & 18  $\pm$1   &  3.6  $\pm$ 0.6 & 150 \\               
14.9705  & 0.609 & 23.23 & 22.79 & 21.99 & 21.64 & 21.30 & 20.97 & *     &  *    & -20.76 &      *  &8.8   $\pm$1.0  &3.38 $\pm$0.45   &2.0  $\pm$0.7   & 23  $\pm$1.5 &  8    $\pm$ 2   & 85 \\   
\hline
22.0274  & 0.504 & 22.19 & 21.91 & 21.05 & 20.70 & 20.40 & 20.35 & 19.90 &   *   & -21.15 & -22.59  & 32.00$\pm$3.0  &11.85$\pm$1.03   & 45$\pm$4.0     & 32  $\pm$5   &  0.67 $\pm$1.24 & 50 \\   
22.0417  & 0.593 & 24.47 & 23.90 & 23.31 & 22.37 & 21.97 & 21.84 & 21.22 &   *   & -19.85 & -21.38  &  6.90$\pm$0.26 &2.86 $\pm$1.4    & 2.60$\pm$0.26  & 8.5 $\pm$1   &  2.51 $\pm$0.49 & 50 \\   
22.0429  & 0.624 & 24.77 & 24.02 & 23.46 & 22.45 & 21.89 & 21.67 & 20.49 &   *   & -20.03 & -22.01  & 10.20$\pm$0.33 &3.98 $\pm$0.46   & 3.60$\pm$0.33  & 31  $\pm$5   &  14.90$\pm$2.38 & 100 \\           
22.0576  & 0.886 & 23.58 & 23.35 & 22.89 & 22.61 & 21.98 & 21.75 &     * &  0.35 & -21.33 &      *  & 31.00$\pm$0.33 &12.22$\pm$0.41   & 11.70$\pm$1.32 & 42  $\pm$4   &  10.37$\pm$1.21 & 105 \\           
22.0599  & 0.887 & 23.28 & 22.99 & 22.57 & 22.22 & 21.68 & 21.53 &     * &  0.04 & -20.79 &      *  & 25.00$\pm$0.33 &11.09$\pm$0.19   & 18.90$\pm$0.66 & 14.5$\pm$1.5 &  1.75 $\pm$0.40 & 200 \\     
22.0770  & 0.816 & 23.80 & 23.67 & 23.22 & 22.75 & 22.04 & 21.96 & 21.71 &   *   & -21.02 & -22.07  & 10.50$\pm$0.26 &2.69 $\pm$1.3    & 4.00$\pm$0.33  & 14  $\pm$2   &  2.16 $\pm$0.69 & 120 \\     
22.0919  & 0.472 & 22.41 & 22.51 & 21.92 & 21.91 & 21.33 & 21.50 & *     &   *   & -19.77 &      *  & 51.00$\pm$0.66 &29.63$\pm$0.75   &143.00$\pm$0.66 & 79.8$\pm$4.0 &  1.0  $\pm$ 2.0 & 230 \\        
22.1119  & 0.514 & 22.21 & 21.89 & 21.20 & 20.40 & 20.03 & 19.80 & *     &   *   & -21.42 &      *  & 20.80$\pm$0.46 &11.55$\pm$2.10   & 10.00$\pm$5.0  & 86  $\pm$5   &  34   $\pm$3    & 100 \\
22.1313  & 0.817 & 23.49 & 23.44 & 23.34 & 22.65 & 21.90 & 21.77 & 21.16 &  0.73 & -21.09 & -22.86  & 13.60$\pm$0.40 &4.89 $\pm$0.9    & 8.20$\pm$0.40  & 12  $\pm$2   &  3.23 $\pm$0.73 & 50 \\     
\enddata
%\tablenotemark{a} In der Tabelle
\tablenotetext{a}{half-light radius $r_{0.5}$}
\tablenotetext{b}{Fluxes in
  $10^{-17}\rm{ergs}\,\rm{s}^{-1}\rm{cm}^{-2}$}
\tablenotetext{c}{from the CFHT optical observations presented in LCS03}
%\tablenotetext{d}{equivalentwidth of \Ha\, in \AA\,}
%\tablerefs{}
%\tablecomments
\end{deluxetable}

%##########################################################################################
%
%                               %DETAILS OBSERVATIONS
%
%###########################################################################################

\clearpage

\begin{table}[h!]
\caption{\label{ObsISAAC} Spectroscopic follow-up
  with VLT and Keck
 }
\begin{center}
\begin{tabular}{c c c c c c}
\hline
\hline
\#   & Filter & $t_{exp}$(s) & Night & Seeing & Telluric Standard\\
\hline
&  &  &VLT-ISAAC  &  &   \\
\hline
03.0085 & SZ  &  2400 & 2003 Nov 18/19 & 1\arcsec.1 &  Hip031190\\
03.0145 & SZ  &  2400 & 2003 Nov 18/19 & 1\arcsec.2 &  Hip031190\\
03.0488 & SZ  &  1800 & 2004 Oct 08/09& 1\arcsec.0 &  Hip026098 \\
03.1309 & SZ  &  2400 & 2003 Dec 03/04 & 1\arcsec.1 &  Hip031190 \\
03.9003 & SZ  &  1800 & 2004 Sep 22/23& 1\arcsec.2 &  Hip023551 \\
22.0274 & SZ &  1200 & 2002 Oct 11/12& 1\arcsec.2 &  Hip109332 \\
22.0417 & SZ &  4800 & 2002 Oct 17/18& 0\arcsec.7 &  Hip109332 \\
22.0429 & SZ &  2400 & 2002 Oct 10/11& 1\arcsec.3 &  Hip109332 \\
22.1119 & SZ &  1800 & 2002 Oct 09/10 & 1\arcsec.2 &  Hip109332 \\
\hline
03.0125 & J   &  2400 & 2003 Nov 18/19 & 0\arcsec.7 &  Hip031190 \\
03.1112 & J   &  2400 & 2003 Dec 03/04 & 0\arcsec.8 &  Hip031190 \\
03.1367 & J   &  2400 & 2003 Dec 03/04 & 0\arcsec.8 &  Hip031190 \\
03.1534 & J   &  2400 & 2004 Oct 08/09& 1\arcsec.0 &  Hip026098 \\
22.0576 & J  &  1200 & 2002 Oct 11/12& 1\arcsec.2 &  Hip109332 \\
22.0599 & J  &  1200 & 2002 Oct 11/12& 1\arcsec.2 &  Hip109332 \\
22.0770 & J  &  3600 & 2002 Oct 10/11& 0\arcsec.9 &  Hip100556 \\
22.1313 & J  &  2400 & 2002 Oct 06/07 & 0\arcsec.7 &  Hip000183 \\
\hline
&  &  &Keck-NIRSPEC\tablenotemark{a}  &  &   \\
\hline
10.1116 & NS2  &  2400 & 2004 Mar 29/30 & 0\arcsec.7 & HD89239   \\
10.2418 & NS2  &  2400 & 2004 Mar 30/31 & 0\arcsec.5 & 86 UMa    \\
14.0072 & NS1  &  2400 & 2004 Mar 30/31 & 0\arcsec.5 & 86 UMa    \\
14.0129 & NS3  &  2400 & 2004 Mar 30/31 & 0\arcsec.7 & HD129653  \\
14.0272 & NS2  &  2400 & 2004 Mar 29/30 & 0\arcsec.7 & HD89239   \\
14.0393 & NS1  &  2400 & 2000 June 14/15& 0\arcsec.7 &  HD89239 \\
14.0985 & NS2  &  2400 & 2004 Mar 29/30 & 0\arcsec.7 & HD129653  \\
14.1258 & NS1  &  2400 & 2004 Mar 30/31 & 0\arcsec.6 & 86 UMa    \\
14.1386 & NS2  &  2400 & 2004 Mar 29/30 & 0\arcsec.7 & HD129653  \\
14.1466 & NS2  &  2400 & 2004 Mar 29/30 & 0\arcsec.7 & HD129653  \\
14.9705 & NS1  &  2400 & 2004 Mar 30/31 & 0\arcsec.6 & HD129653  \\
\hline
\end{tabular}
\end{center}
\tablenotetext{a}{NIRSPEC uses a set of
  order-sorting filters to cover the range from 0.95 to 2.6 microns;
  these are  called ``NIRSPEC-1'', ``NIRSPEC-2'', etc., and
  abbreviated in the table   as ``NS1''($0.947-1.121 \mu m$),
  ``NS2''($1.089-1.293 \mu m$), and ``NS3''($1.143-1.375 \mu m$).}
\end{table}

%################################################################################
%
%               The MODEL based on KD02
%
%################################################################################
\clearpage

\begin{table}[h]
\caption{\label{modelgrid} The model grid used to calculate  $A_{V}$,$q$, and [O/H]}  
\begin{center}
\begin{tabular}{c c}
\hline
Parameter & Range   \\
\hline
$A_{V}$ (extinction parameter) & $0< A_{V} < 3.01$ in 302 steps \\
$q$ (ionisation parameter) & $5 \cdot 10^{6}$cm\,s$^{-1}< q < 300 \cdot 10^{6}$cm\,s$^{-1}$ in 201 steps \\
O/H (oxygen abundance) & $7.6< O/H < 9.4$ in 201 steps \\
\hline
\end{tabular}
\end{center}
\end{table}

%################################################################################
%
%               DERIVED QUANTITIES
%
%################################################################################
\clearpage

\begin{table}[h!]
\caption{\label{CFRSHaNII2} Derived quantities for the 30 CFRS galaxies   }
\begin{center}
\begin{tabular}{c c c c c c  c}
\hline
\hline
\#      &  $z$  &     \Ha/\Hb   &   [NII]/\Ha  & A$_{V}$\tablenotemark{a}  & SFR  &[O/H]\tablenotemark{a}\\
\hline
\hline
03.0085 & 0.609 & 3.50$\pm$1.59 & 0.09$\pm$0.08  & 1.34$^{+0.40}_{-1.34}$  &   5.08$^{+ 2.53}_{- 3.37}$ & 8.36$^{+0.28}_{-0.55}$ \\
03.0125 & 0.789 & 2.89$\pm$0.93 & 0.26$\pm$0.07  & 1.57$^{+0.23}_{-0.25}$  &   9.86$^{+ 3.42}_{- 2.79}$ & 8.62$^{+0.10}_{-0.09}$ \\
03.0145 & 0.603 & 3.54$\pm$0.57 & 0.19$\pm$0.06  & 0.20$^{+1.34}_{-0.20}$  &   3.02$^{+ 5.92}_{- 0.76}$ & 8.90$^{+0.04}_{-0.09}$ \\
03.0488 & 0.605 & 3.31$\pm$0.50 & 0.25$\pm$0.06  & 0.36$^{+0.49}_{-0.36}$  &   4.65$^{+ 2.61}_{- 1.41}$ & 8.88$^{+0.05}_{-0.08}$ \\
03.0599 & 0.479 & 5.11$\pm$0.99 & 0.13$\pm$0.09  & 0.96$^{+0.23}_{-0.25}$  &   2.89$^{+ 0.68}_{- 0.59}$ & 8.57$^{+0.13}_{-0.14}$ \\
03.1112 & 0.768 & 3.05$\pm$1.49 & 0.22$\pm$0.10  & 0.15$^{+0.68}_{-0.15}$  &   1.91$^{+ 1.97}_{- 0.62}$ & 8.78$^{+0.08}_{-0.21}$ \\
03.1309 & 0.617 & 4.54$\pm$0.80 & 0.55$\pm$0.07  & 1.36$^{+0.49}_{-0.43}$  &  19.59$^{+10.58}_{- 6.41}$ & 9.01$^{+0.05}_{-0.06}$ \\
03.1367 & 0.703 & 5.84$\pm$0.67 & 0.31$\pm$0.03  & 2.08$^{+0.38}_{-0.31}$  &  25.99$^{+11.12}_{- 7.00}$ & 8.87$^{+0.05}_{-0.06}$ \\
03.1534 & 0.794 & 3.32$\pm$0.40 & 0.05$\pm$0.02  & 0.45$^{+0.26}_{-0.33}$  &   6.12$^{+ 2.03}_{- 1.80}$ & 8.25$^{+0.09}_{-0.11}$ \\
03.9003 & 0.618 & 8.42$\pm$1.68 & 0.27$\pm$0.04  & 1.31$^{+0.31}_{-0.28}$  &  27.36$^{+10.00}_{- 7.07}$ & 9.14$^{+0.02}_{-0.02}$ \\
\hline
10.1116 & 0.709 & 4.47$\pm$1.72 & 0.45$\pm$0.07  & 1.27$^{+0.61}_{-0.94}$  &   5.21$^{+ 3.52}_{- 2.77}$ & 8.93$^{+0.12}_{-0.14}$ \\ 
10.2418 & 0.796 & 6.15$\pm$0.82 & 0.44$\pm$0.05  & 1.57$^{+0.40}_{-0.33}$  &  24.48$^{+10.98}_{- 6.85}$ & 8.99$^{+0.04}_{-0.05}$ \\  
\hline
14.0072 & 0.621 & 4.31$\pm$0.42 & 0.27$\pm$0.04  & 1.21$^{+0.29}_{-0.28}$  &   5.52$^{+ 1.83}_{- 1.37}$ & 8.78$^{+0.05}_{-0.06}$ \\  
14.0129 & 0.903 & 4.05$\pm$1.24 & 0.37$\pm$0.05  & 0.93$^{+1.11}_{-0.76}$  &   7.65$^{+10.64}_{- 3.57}$ & 9.01$^{+0.09}_{-0.14}$ \\  
14.0272 & 0.670 & 3.53$\pm$0.67 & 0.44$\pm$0.09  & 0.58$^{+0.47}_{-0.46}$  &   6.28$^{+ 3.48}_{- 2.26}$ & 9.16$^{+0.04}_{-0.03}$ \\ 
14.0393 & 0.603 & 3.44$\pm$0.45 & 0.28$\pm$0.09  & 0.33$^{+0.44}_{-0.33}$  &   6.92$^{+ 3.18}_{- 1.81}$ & 9.00$^{+0.04}_{-0.04}$ \\
14.0985 & 0.809 & 1.70$\pm$0.46 & 0.42$\pm$0.21  & 0.00$^{+0.17}_{-0.00}$  &   1.17$^{+ 0.29}_{- 0.12}$ & 8.71$^{+0.09}_{-0.21}$ \\ 
14.1258 & 0.647 & 3.06$\pm$0.21 & 0.09$\pm$0.06  & 0.21$^{+0.45}_{-0.21}$  &   2.82$^{+ 1.30}_{- 0.53}$ & 7.87$^{+0.11}_{-0.08}$ \\  
14.1386 & 0.744 & 6.75$\pm$1.09 & 0.39$\pm$0.06  & 3.01$^{+0.03}_{-0.08}$  &  70.02$^{+ 8.50}_{-10.34}$ & 8.54$^{+0.04}_{-0.06}$ \\ 
14.1466 & 0.674 & 5.61$\pm$0.57 & 0.20$\pm$0.04  & 1.41$^{+0.22}_{-0.21}$  &   7.73$^{+ 1.82}_{- 1.45}$ & 8.44$^{+0.06}_{-0.05}$ \\ 
14.9705 & 0.609 & 6.80$\pm$1.01 & 0.35$\pm$0.09  & 2.46$^{+0.26}_{-0.32}$  &  16.26$^{+ 4.60}_{- 4.17}$ & 8.75$^{+0.11}_{-0.10}$ \\  
\hline
22.0274 & 0.504 & 2.70$\pm$0.48 & 0.02$\pm$0.04  & 0.00$^{+0.16}_{-0.00}$  &   2.47$^{+ 0.73}_{- 0.39}$ & 8.42$^{+0.14}_{-0.13}$ \\
22.0417 & 0.593 & 2.97$\pm$1.50 & 0.30$\pm$0.07  & 1.18$^{+0.37}_{-1.18}$  &   2.26$^{+ 1.03}_{- 1.40}$ & 8.78$^{+0.22}_{-0.11}$ \\  
22.0429 & 0.624 & 7.79$\pm$1.55 & 0.48$\pm$0.11  & 2.62$^{+0.34}_{-0.40}$  &  26.06$^{+12.53}_{- 9.64}$ & 8.84$^{+0.07}_{-0.07}$ \\  
22.0576 & 0.886 & 3.44$\pm$0.35 & 0.25$\pm$0.04  & 0.01$^{+0.30}_{-0.01}$  &  12.91$^{+ 4.61}_{- 1.31}$ & 8.97$^{+0.02}_{-0.03}$ \\
22.0599 & 0.887 & 1.31$\pm$0.14 & 0.12$\pm$0.03  & 0.00$^{+0.05}_{-0.00}$  &   4.44$^{+ 0.64}_{- 0.46}$ & 8.51$^{+0.08}_{-0.09}$ \\
22.0770 & 0.816 & 5.20$\pm$2.62 & 0.15$\pm$0.05  & 1.29$^{+0.27}_{-0.31}$  &   8.77$^{+ 3.39}_{- 2.75}$ & 8.56$^{+0.14}_{-0.11}$ \\  
22.0919 & 0.472 & 2.69$\pm$0.15 & 0.01$\pm$0.03  & 0.00$^{+0.15}_{-0.00}$  &   5.26$^{+ 0.89}_{- 0.26}$ & 8.38$^{+0.17}_{-0.11}$ \\ 
22.1119 & 0.514 & 7.45$\pm$1.42 & 0.40$\pm$0.04  & 2.73$^{+0.30}_{-0.59}$  &  48.92$^{+15.23}_{-18.70}$ & 8.89$^{+0.08}_{-0.09}$ \\
22.1313 & 0.817 & 2.45$\pm$0.61 & 0.27$\pm$0.08  & 0.00$^{+0.44}_{-0.00}$  &   3.00$^{+ 1.79}_{- 0.50}$ & 8.86$^{+0.04}_{-0.08}$ \\  
\hline                                                
\end{tabular}                                         
\end{center}                                          
\tablenotetext{a}{The quoted uncertainties are the purely statistical
  measurement uncertainties propagating through to parameter
  determinations. These are addressed by our $\chi^{2}$ analysis. They
  reflect both the quality of the data and the gradients (and
  degeneracies) in the KD02 models. For more details see Section \ref{erbars}.}
\end{table}


\clearpage


\begin{thebibliography}{}



\bibitem[Brocklehurst(1971)]{brockle} Brocklehurst M.  1971 MNRAS, 153, 471

  
\bibitem[Carollo and Lilly(2001, hereafter CL01)]{calilly01pap} Carollo, C.
  M. \& Lilly, S. J. 2001, ApJ, 548, 153, CL01



\bibitem[Fioc \& Rocca-Volmerange(1999)]{fiocrocca99} Fioc, M. \& Rocca-Volmerange, B., 1999, astro-ph/9912179

\bibitem[Hammer et al.(2001)]{hammer01} Hammer, F., Gruel, N., Thuan,
  T. X., Flores, H., and Infante, L., 2001, ApJ, 550, 570


\bibitem[Hippelein et al.(2003)]{hippelein} Hippelein, H., Maier, C., Meisenheimer, K.  et
 al.  2003, A\&A, 402, 65

\bibitem[Horne(1986)]{horne} Horne K. 1986,  PASP, 98, 609

\bibitem[Izotov et al.(1994)]{izotov} Izotov, Y.~I., Thuan, T.~X. \&
  Lipovetsky, V.~A. 1994,  ApJ, 435, 647


\bibitem[Jansen et al.(2000)]{jansen} Jansen, R.~A., Fabricant, D., Franx, M. et al,
  2000, ApJS, 126, 331

\bibitem[Kennicutt(1998)]{ken98} Kennicutt, R. C., Jr. 1998, ARA\&A, 36, 189

\bibitem[Kewley et al.(2001)]{kewley01} Kewley, L.J., Heisler, C.A.,  Dopita, M.A.\&
  Lumsden, S.  2001, ApJS, 132, 37


\bibitem[Kewley \& Dopita(2002)]{kewdop02pap} Kewley, L.J. \& Dopita, M.A., 2002, ApJSS, 142, 35, KD02

\bibitem[Kobulnicky et al.(1999)]{kob99} Kobulnicky, H. A.,  Kennicutt. R.C., and Pizgano, J. L., 1999, ApJ, 514, 544

\bibitem[Kobulnicky et al.(2003)]{kob03} Kobulnicky, H. A., Wilmer C. N. A., Weiner, B. J. et al.  2003, ApJ, 599, 1006

\bibitem[Kobulnicky \& Kewley(2004)]{kobkew04} Kobulnicky, H. A. \&
  Kewley, L.J.  2004, ApJ 617, 240


\bibitem[Lamareille et al.(2004)]{lamar04} Lamareille, F., Mouhcine,
  M., Contini, T., Lewis, I. \& Maddox, S., 2004, MNRAS, 350, 396

\bibitem[Lilly et al.(2003, hereafter LCS03)]{lilly03} Lilly, S.J, Carollo, C.M. \&
  Stockton, A. 2003, ApJ, 597, 730, LCS03


\bibitem[Maier et al.(2004)]{maier04} Maier, C., Meisenheimer, K., Hippelein, H.,
  2004, A\&A, 418, 475

\bibitem[Maier et al.(2005)]{maier05} Maier, C. et al.,
  2005, in preparation


\bibitem[Mehlert et al.(2002)]{mehlert02} Mehlert, D., Noll, S., Appenzeller, I. et al. 2002, A\&A, 393, 809

\bibitem[Melbourne \& Salzer(2002)]{melbsal} Melbourne, J. \& Salzer, J. J. 2002, AJ, 123, 2302

\bibitem[Osterbrock(1989)]{osterb} Osterbrock, D.E. 1989, Astrophysics of Gaseous Nebulae and Active
Galactic Nuclei (Mill Valley: University Science Books)


\bibitem[Pagel et al.(1979)]{pagel79}  Pagel, B. E. J., Edmunds, M. G., Blackwell, D. E. et al. 1979, MNRAS, 189, 95



\bibitem[Pettini et al.(2002)]{pettini02} Pettini, M., Rix, S.A., Steidel, C.C.  et al. 2002, Ap\&SS, 281, 461

\bibitem[Pettini et al.(2003)]{pettini03}  Pettini, M. 2003, astro-ph/0303272

\bibitem[Pettini \& Pagel(2004)]{petpag} Pettini, M. \& Pagel, B.E.J. MNRAS, 2004, 348, 59

\bibitem[Salzer et al.(2005)]{salzer05} Salzer, J. J., Lee, J.C.,
  Melbourne, J. et al., 2005, ApJ, 624, 661


\bibitem[Seaton(1979)]{seaton} Seaton, M. J. MNRAS 1979, 187, 73

\bibitem[Shapley et al.(2004)]{shapley04} Shapley, A., Erb, D.~K.,
  Pettini, M. et al.  2004,  ApJ, 612, 108

\bibitem[Tremonti et al.(2004)]{tremon04} Tremonti, C.~A., Heckman,
  T.~M., Kauffmann, G. et al. 2004, ApJ, 613, 898 


\bibitem[van Dokkum(2001)]{vDo01} van Dokkum, P. G. 2001, \pasp, 113, 1420



\bibitem[Whitford(1958)]{whitford}  Whitford, A. E., 1958, AJ,  63, 201

\bibitem[Wolf et al.(2003)]{wolf03} Wolf, C., Meisenheimer, K., Rix, H.-W.  et al 2003, A\&A, 401, 73 

\end{thebibliography}
\end{document}